\def\draftversion{false}
\RequirePackage{ifthen}

\ifthenelse{\equal{\draftversion}{true}}{
  \documentclass[aps,prb,showpacs,longbibliography,galley,
                 superscriptaddress]{revtex4}
}{
  \documentclass[aps,prb,showpacs,longbibliography,twocolumn,
                 superscriptaddress]{revtex4-1}
}

\usepackage{amsmath,graphicx,latexsym,bm}

\usepackage{soul}  

\usepackage[usenames,dvipsnames]{color}

\newcommand{\red}{\textcolor{Red}}

\ifthenelse{\equal{\draftversion}{true}}{
  \marginparwidth 2.7in
  \marginparsep 0.5in
  \newcounter{comm} 
  \def\commnext{\stepcounter{comm}}
  \def\commtext{{\bf\color{blue}[\arabic{comm}]}}
  \def\commmar{{\bf\color{blue}[\arabic{comm}]}}
  \def\dvm#1{\commnext\marginpar{\small DV\commmar: #1}\commtext}
  \def\dvr#1{\dvm{\red{#1}}}
  \def\msm#1{\commnext\marginpar{\small MS\commmar: #1}\commtext}
  \def\mlab#1{\marginpar{\small\bf #1}}

}{
  \def\dvm#1{}
  \def\dvr#1{}
  \def\msm#1{}
  \def\mlab#1{}

}

\newcommand{\mylabel}[1]{\label{#1}} 

\newcommand{\cur}[1]{\tilde{#1}} 
\newcommand{\cop}[1]{\hat{\tilde{#1}}} 

\begin{document}

\title{Quantum theory of mechanical deformations}

\author{Massimiliano Stengel}
\affiliation{ICREA - Instituci\'o Catalana de Recerca i Estudis Avan\c{c}ats, 08010 Barcelona, Spain}
\affiliation{Institut de Ci\`encia de Materials de Barcelona
(ICMAB-CSIC), Campus UAB, 08193 Bellaterra, Spain}

\author{David Vanderbilt}
\affiliation{Department of Physics and Astronomy, Rutgers University, Piscataway, New Jersey 08854-8019, USA}

\date{\today}

\begin{abstract}
We construct a general metric-tensor framework for treating
inhomogenous adiabatic deformations applied to crystalline insulators,
by deriving an effective time-dependent Schr\"odinger equation in the
undistorted frame.
The response can be decomposed into ``static'' and ``dynamic'' terms
that correspond, respectively, to the amplitude and the velocity of the
distortion.
We then focus on the dynamic contributon,
which takes the form of a gauge field entering the effective
Hamiltonian, in the linear-response limit.
We uncover an intimate
relation between the dynamic response to the rotational component of
the inhomogeneous deformation and the diamagnetic response to
a corresponding inhomogeneous magnetic field.
We apply this
formalism to the theory of flexoelectric response, where we
resolve a previous puzzle by showing that the currents generated by
the dynamic term, while real, generate no bound charges even at
surfaces, and so may be dropped from a practical theory of
flexoectricity.
\end{abstract}

\maketitle

\section{Introduction}

Mechanical deformations are among the most basic perturbations that can
be applied to a crystalline solid, and their response is at the origin
of many basic materials functionalities, such as elasticity or
piezoelectricity.
The development of theoretical approaches to calculate and predict these
properties from first principles has marked notable milestones for
modern electronic-structure theory, paralleling the equally important
development of density-functional perturbative approaches to lattice
dynamics.
In the case of uniform deformations, methods to compute the relevant
response coefficients are now well established, and part of the most
popular simulation packages that are available to the public.
Yet, with the rising interest in flexoelectricity, and more generally in
functionalities that depend on the \emph{gradient} of the strain field
rather than on the strain itself, the existing computational approaches
are of limited applicability, and their generalization to cases where
the deformation is inhomogeneous appears far from obvious.

Flexoelectricity, describing the polarization response of a crystalline
insulator to a strain gradient, has received considerable attention in
the past few years because of its fundamental interest and potential
relevance to energy and information technologies.
Recent advances in first-principles methods have given a considerable
boost to the field.
The theoretical understanding of flexoelectric phenomena, however, still
presents daunting conceptual and practical challenges, many of which are
still unresolved.
The purely electronic (clamped-ion) contribution to the flexoelectric
response, for example, is riddled with subtleties, and proper
methodologies to compute it in the most general case are still missing.
(Lattice-mediated effects are comparatively much simpler to understand,
both conceptually and computationally -- they consists in the dynamical
dipoles produced by the internal relaxations of the primitive cell, and
bear many analogies to the point-charge model proposed by
Tagantsev~\cite{Tagantsev-86} long ago.)

The main issue resides in that, in order to define the transverse
components of the clamped-ion flexoelectric tensor at the bulk level,
one needs, in principle, to access the \emph{microscopic} polarization
response to a variety of lattice distortions (e.g., long-wavelength
acoustic phonons, or displacements of an isolated atom).
Indeed, treatments based on the Berry-phase formula are ruled out
because a strain gradient breaks translational periodicity;
charge-density based theories are not viable either, as they yield only
partial information on the flexoelectric tensor components.
Calculating the microscopic polarization response implies establishing a
time-dependent perturbative framework, where the quantum-mechnical
probability current is monitored in the course of a slow distortion of
the crystal.
Such a procedure, however, falls outside the capabilities of the
publically available electronic-structure packages. An
implementation of the current-density based theory of
flexoelectricity has only very recently been presented in 
Ref.~\onlinecite{cyrus}.
This implementation required the resolution of some challenging
formal issues regarding the current-density response to a
macroscopic deformation.  A brief account of those issues was given
in Ref.~\onlinecite{cyrus}, but are described more thoroughly and 
in greater depth in the present manuscript.

%

The first, obvious, question concerns the physical representation of a
microscopic observable, such as the electronic probability current, in a
context where the boundary conditions of the Hamiltonian change in the
course of the transformation.
In a nutshell, even if we limit our attention to the simplest case of a
uniform strain (these issues become all the more severe if the
deformation is inhomogeneous), the atomic distortion pattern that one
needs, in principle, to apply in order to strain the crystal grows
linearly with the distance from the origin. (Recall that the macroscopic
strain is related to the first gradient of the displacement field.)
This has two undesirable consequences: (i) the perturbation (and hence
all the microscopic response functions associated to it) is nonperiodic
and origin-dependent, even if both the initial and the final state enjoy
translational periodicity; (ii) the perturbation is never small at the
boundaries of a large crystallite, which complicates its treatment
within linear-response theory.
In Ref.~\onlinecite{artgr} the above problem was elegantly solved by
combining the macroscopic deformation with a simultaneous coordinate
transformation.
This way, one can encode the strain field as a change in the metric of
space while the atoms remain at their original locations, thereby
removing the need for a nonperiodic and unbound lattice distortion.
Also, the transformed coordinate system naturally leads to a sound
definition of microscopic response functions, such as polarization,
charge density and local electric fields.

The second conceptual issue is even more subtle, and consists in making
sure that the fundamental response quantities of interest (e.g. the
flexoelectric polarization) are well defined, i.e., that they are
independent of the rotations or translations that were applied to the
crystal in the course of the deformation.
This is required by a proper~\cite{vanderbilt:2000} theory of
electromechanical phenomena, which should depend on physically
meaningful changes in the relative distances between neighboring
material points, and not on their absolute position with respect to some
arbitrary coordinate frame.
In order to make the problem tractable, in
Refs.~\onlinecite{artlin,artgr} we had to make some simplifying
assumptions on how the electronic currents, ${\bf J}({\bf r})$,
respond to a rigid rotation or translation
of the crystal lattice, by postulating that
$${\bf J}({\bf r}) = {\bf v}({\bf r}) \rho({\bf r}),$$
where $\rho$ is the charge density and ${\bf v}$ is the velocity
of the material point ${\bf r}$ imposed by the rototranslation.
This is akin to assuming that the electronic cloud behaves
as a classical charge distribution that is equal to the
true quantum-mechanical one.
As we shall see, this is indeed correct in the case of translations, but
not in the case of rotations, where there is a further contribution
to the current that we neglected in earlier works. It turns out, however,
that this additional piece is curl-free, so it is unclear whether it
affects the results.
Settling this point appears as a clear priority: the theory of
flexoelectricity, as it stands, crucially relies on this assumption in
order to define~\cite{artgr} and calculate~\cite{artcalc} the transverse
components of the bulk flexoelectric tensor.
A fundamental theoretical framework, where the microscopic polarization
currents are derived within a proper quantum-mechanical treatment of
deformations, is needed in order to firm up the results obtained so far,
and thereby pave the way towards future developments in the field.

Here we attack this problem from its very root, by incorporating
coordinate transformations directly into the time-dependent
Schr\"odinger equation. This allows us to perform a formal analysis of
the electronic probability current that develops in the course of an
arbitrary mechanical deformation, and thereby to identify the relevant
physical contribution to the polarization response in the most general
case.
Interestingly, we find that a non-uniform
deformation is generally accompanied by
``gauge currents'' produced by local rotations of the sample.
These currents are divergenceless and correspond to
the circulating diamagnetic currents generated by an applied magnetic
field (${\bf B}$).
This result is explained heuristically by recalling Larmor's theorem, which
relates the Lorentz force on a charged particle in a uniform ${\bf B}$-field
to the Coriolis force on a massive object in a uniformly rotating frame.
By performing a long-wavelength analysis in the limit of small
deformations, we demonstrate that the bulk flexoelectric tensor
has a contribution from these gauge fields that is proportional to
the bulk diamagnetic susceptibility of the material.
Remarkably, such a contribution is exactly cancelled by an equal and
opposite surface term. One is therefore free to remove this term from
both sides, leaving a description of the flexoelectric polarization that
is consistent with the charge-density-based strategy of
Ref.~\onlinecite{artcalc}.

The present results demonstrate, once more, the intimate connection
between surface and bulk contributions to the flexoelectric effect, and
the intriguing connections between the latter phenomenon and other,
apparently unrelated, areas of research (in this case, orbital
magnetism).
In addition to providing a firm foundation to the existing theory of
flexoelectricity, we also provide an explicit derivation of how a
generalized (and time-dependent) coordinate transformation of space is
reflected in the most basic quantum-mechanical operators, such as the
Hamiltonian or the probability current. This can be of immediate
usefulness to a wide range of physical problems, within and
beyond~\cite{ortix} the specific context of this work.

\section{General theory}

We shall consider a generic time-dependent deformation of the crystal
lattice, where all atoms move from their original location,
${\bf R}_{l\kappa}^0$, according to a continuous vector function of space
and time, ${\bf r}(\bm{\xi},t)$,
\begin{equation}
{\bf R}_{l\kappa}(t) = {\bf r} ({\bf R}_{l\kappa}^0,t).
\label{clamped}
\end{equation}
(Recall that $\kappa$ and $l$ are sublattice and cell
indices, respectively.)
The physical effects of the deformation described by ${\bf
r}(\bm{\xi},t)$ are best treated by operating an analogous coordinate
transformation that brings every atom back to its original
position.~\cite{artgr}
This means that the atoms are immobile in the (generally curvilinear)
$\bm{\xi}$-frame, but the frame itself evolves with respect to the
Cartesian laboratory frame.
All the effects of the mechanical perturbation are, in other words,
encoded in the metric of the deformation, rather than in an atomic
displacement pattern.
Note that describing deformations in terms of coordinate transformations 
in a quantum mechanical context is a strategy that already has a relatively long 
history~\cite{gygi-93,hamann-metric,nussinov-03,nussinov-16}; our
approach has several points of contact with these works.

To see how the metric change affects the electronic Hamiltonian, it is
useful to introduce a number of auxiliary quantities that will come
handy later in the derivation.
%
The first is the so-called deformation gradient,
\begin{equation}
h_{i \alpha} = \frac{\partial r_i}{\partial \xi_\alpha}.
\end{equation}
The determinant of the deformation gradient, $h=\det ({\bf h})$, gives
the local volume change with respect to the unperturbed lattice
configuration.
From the deformation gradient we can construct the metric tensor,
%
%
\begin{equation}
g_{\alpha\beta} = \frac{\partial r_i}{\partial \xi_\alpha} \frac{\partial r_i}{\partial \xi_\beta} = h_{i \alpha} h_{i \beta}
 = ( {\bf h}^{\rm T} {\bf h})_{\alpha\beta},
\mylabel{gdef}
\end{equation}
which is another central quantity of the formalism;
its determinant is $g= \det ({\bf g}) = h^2$.
(Here and in the following we us an implicit sum notation on
indices, with Roman and Greek indices used for Cartesian and
curvilinear frames respectively.)

We shall define the wavefunctions in the deformed space in such a way
that they comply with the basic orthonormality requirements.
This means writing
\begin{equation}
\psi({\bf r}) = h^{-1/2}\, \cur{\psi} (\bm{\xi}).
\mylabel{psi-renorm}
\end{equation}
It is easy to show that the wavefunctions $\psi$ are orthonormal in the
Cartesian space provided that the ``curvilinear'' wavefunctions
$\cur{\psi}$ are orthonormal in the $\bm{\xi}$-space,
\begin{eqnarray}
\int d^3 r \, \psi^*_m ({\bf r}) \psi_n ({\bf r}) &=&
\int d^3 r \, h^{-1}\, \cur{\psi}^*_m (\bm{\xi}({\bf r})) \cur{\psi}_n (\bm{\xi}({\bf r})) \nonumber \\
&=& \int d^3 \xi \, \cur{\psi}^*_m (\bm{\xi}) \cur{\psi}_n (\bm{\xi}).
\end{eqnarray}

Note that we shall work in a time-dependent context, which is necessary in
order to be able to discuss the polarization response. In doing so we assume
\begin{equation}
\psi({\bf r},t) = h^{-1/2}\, \cur{\psi} (\bm{\xi},t),
\mylabel{psi-renorm-t}
\end{equation}
i.e., the phase evolution of $\cur{\psi}$ is locked to that of
$\psi$.
The choice of the phase relation between $\cur{\psi}$ and
$\psi$ is mostly a matter of convention, and can be regarded as
a ``gauge freedom'' of the transformed wavefunctions.
Indeed, one could postulate
$\psi({\bf r},t) = e^{i\varphi} h^{-1/2}\, \cur{\psi} (\bm{\xi},t),$
where $\varphi$ is an arbitrary function of space and time.
While the physical conclusions should not depend on $\varphi$,
the specific form of the time-dependent Schr\"{o}dinger equation
in the comoving frame does. In particular, unfamiliar terms may
arise in the Hamiltonian whose physical interpretation needs some caution;
we shall briefly discuss an illustrative example in Sec.~\ref{sec:gal}

In the following Sections our goal will be to start from a conventional
Schr\"odinger equation, written in Cartesian ${\bf r}$-space, and
progressively work out the curvilinear version in $\bm{\xi}$-space,
where the electronic wavefunctions are described by $\cur{\psi}$.
%
%
%

\subsection{Time-dependent Schr\"odinger equation}

\mylabel{sec:tdse}

The time-dependent Schr\"odinger equation can be written in the original
Cartesian frame as
\begin{equation}
i \frac{\partial}{\partial t} \psi ({\bf r},t) = \left[-\frac{\nabla^2}{2} + V({\bf r},t) \right] \psi ({\bf r},t),
\mylabel{schr-cart}
\end{equation}
where we have set $\hbar=m_e=1$.
Multiplying through by $\sqrt{h}$ and carrying out the
coordinate transformations, this becomes (the detailed derivations can be found
in the Appendix)
\begin{equation}
i\frac{\partial}{\partial t} \; \cur{\psi} = \cop{\mathcal{H}} \cur{\psi} ,
\mylabel{schr-new}
\end{equation}
where the new effective Hamiltonian operator,
\begin{eqnarray}
\cop{\mathcal{H}} &=&
  \frac{1}{2} (\cop{p}_\beta - A_\beta ) g^{\beta \gamma}
        (\cop{p}_\gamma - A_\gamma ) \nonumber\\
 &&\qquad + \cur{V} + V_{\rm geom}
        - \frac{1}{2} \phi,
\mylabel{schr-xi}
\end{eqnarray}
contains contributions arising not only from the potential and kinetic terms on the
right-hand side of Eq.~(\ref{schr-cart}), but also from the
time-derivative term on the left.
Here $\cop{p}_\beta=-i \partial/\partial \xi_\beta$ indicates the
canonical momentum in curvilinear space, $g^{\beta \gamma}= ({\bf g}^{-1})_{\beta \gamma}$
is the inverse metric tensor, $\cur{V}({\bm{\xi}},t) = V({\bf r}({\bm{\xi}},t),t)$ is the
external potential represented in the curvilinear frame, and we have introduced a number of
additional quantities.
First, the ``geometric'' scalar potential $V_{\rm geom}$ originates from the
kinetic energy operator, and reads as
\begin{eqnarray}
V_{\rm geom} &=& \frac{1}{2} \mathcal{A}_\beta g^{\beta \gamma} \mathcal{A}_\gamma +
\frac{1}{2} \partial_\beta (g^{\beta \gamma} \mathcal{A}_\gamma), \mylabel{vgeom}\\
\mathcal{A}_\alpha  &=& \frac{1}{2 h} \frac{\partial h}{\partial \xi_\alpha} = \frac{1}{2} \frac{\partial \ln (h)}{\partial \xi_\alpha}.
\mylabel{mcA-def}
\end{eqnarray}
(Note the close relationship of the auxiliary field $\mathcal{A}_\alpha$ to the contracted Christoffel
symbol.)
Second, we have a further scalar and vector potential field originating from
the time derivative,
\begin{eqnarray}
\phi &=& 
\frac{ \partial r_i }{\partial t}
  \frac{ \partial r_i }{\partial t}, \mylabel{phidef} \\
   A_\gamma &=& \frac{ \partial r_i }{\partial \xi_\gamma}
    \frac{ \partial r_i }{\partial t} \Big|_{\bm{\xi}}.
\mylabel{Adef}
\end{eqnarray}
Interestingly, both ${\bf A}$ and $\phi$ have the same form as the
metric tensor elements in Eq.~(\ref{gdef}),
except that one or both real-space indices have been replaced here with time.
Eq.~(\ref{schr-xi}), together with definitions (\ref{vgeom}), (\ref{mcA-def}),
(\ref{phidef}) and (\ref{Adef}), constitutes one of our central results.

The present theory of deformations bears an intriguing
similarity to electromagnetism, as in both cases the electronic Hamiltonian
acquires a gauge-dependent
vector and scalar potential contribution
[see the discussion following Eq.~(\ref{psi-renorm-t})].
We shall see in the following that both ${\bf A}$ and $\phi$ have
classical counterparts in the fictitious forces that appear in the
noninertial frame defined by the coordinate transformation.

\subsection{Physical interpretation}

To see the physical interpretation of the new terms appearing in the
Schr\"odinger equation, it is useful to work out a couple of simple
examples.

\subsubsection{Galilean transformations}
\label{sec:gal}

Consider a transformation of the type
\begin{equation}
{\bf r} = \bm{\xi} + {\bf v} t,
\mylabel{gtr}
\end{equation}
where ${\bf v}$ is a vector constant with the dimension of a velocity.
We have $h_{\alpha \beta} = g_{\alpha \beta} = \delta_{\alpha \beta}$, $\mathcal{A}_\alpha = 0$ and $A_\alpha = v_\alpha$.
The result is
\begin{equation}
  i \hbar \frac{\partial \cur{\psi}}{\partial t} = \frac{1}{2 m}
 (\cur{\bf p} - m{\bf v} ) \cdot  (\cur{\bf p} - m{\bf v} )  \cur{\psi} + \left( V - \frac{1}{2} mv^2 \right) \cur{\psi},
\mylabel{gal}
\end{equation}
where we have reintroduced the factors of electron mass, $m$, and
$\hbar$ to better illustrate the physical meaning of the various terms.
We have also used the fact that the potential in the co-moving frame
(i.e., we assume here that the crystal is uniformly moving with respect
to the laboratory frame with the same velocity ${\bf v}$) is independent
of time and equal to the potential of the lattice at rest.

To see that Eq.~(\ref{gal}) is reasonable, consider a free particle
$\psi({\bf r},t)=e^{i{\bf q}_0\cdot{\bf r}} e^{-i\omega_0 t}$
with $\hbar\omega_0=p_0^2/2m$ and ${\bf p}_0=\hbar {\bf q}_0$
in the original frame.
Classically, the particle has momentum ${\bf p}_0-m{\bf v}$
as seen from the moving frame.
Eq.~(\ref{psi-renorm-t}) gives its transformed wavefunction to be
$\cur{\psi}(\bm{\xi},t) =
e^{i{\bf q}_0\cdot\bm{\xi}}
e^{-i(\omega_0-{\bf q}_0\cdot{\bf v})t}$, which is easily verified
to satisfy Eq.~(\ref{gal}).  The first term of Eq.(\ref{gal})
is just the kinetic energy
$({\bf p}_0-m{\bf v})^2/2m$ as seen from the comoving frame; the
extra $-mv^2/2$ term is, however, problematic to the extent that
it implies that the energy in the transformed frame cannot
be associated with the expectation value of the Hamiltonian
operator. (Note that the particle velocity is correct, even
if the transformed wavefunctions appears to have the ``wrong''
phase at first sight.)

The fact that the curvilinear-coordinate Hamiltonian does not
reproduce the correct kinetic energy in the co-moving frame may appear
at first sight as a serious limitation of the present theory.
To ensure that this is not a real issue in the context of this work,
some additional words of comments are in order.
First, note that the Galilean
covariance of the Schr\"odinger equation is
not automatic, but requires a specific assumption about the phase of the
transformed wavefunction.
We could have certainly used such a prescription in Eq.~(\ref{psi-renorm-t}),
and this would have restored the standard form of the Schr\"odinger equation
in the uniformly moving frame.
However, this would have been of little help in the context of more general
displacement fields (e.g. nonuniform in time and/or space); in such cases
it is not possible to reabsorb the new gauge potentials with a phase shift.
Second, solving these issues is not essential to
the scope of this work.
As we shall see shortly, we shall either be concerned with the \emph{static}
energy of the system, or with the dynamical evolution of the wavefunctions
up to first order in the velocity; neither of these is affected by the
spurious $\mathcal{O}(v^2)$ term that stems from the ``dynamic scalar potential''
$\phi$.
Further delving into these intriguing fundamental issues, while desirable
in a general context, would bring us far from our present focus, and therefore
we regard this as a stimulating subject for
future investigation.

\subsubsection{Rotating frame}

\label{sec:larmor}

Consider now a transformation of the type
\begin{equation}
{\bf r} = {\bf R}(t)  \bm{\xi},
\end{equation}
where ${\bf R}(t)$ is a $3 \times 3$ matrix describing a rotation about
a given axis $\hat{\bm{\theta}}$.
We have
\begin{equation}
h_{\alpha \beta} = R_{\alpha \beta}, \qquad h=1,
\end{equation}
which implies that $g_{ij} = g^{ij} = \delta_{ij}$, and that
$\mathcal{A}_j = 0$.
On the other hand, we have
\begin{equation}
{\bf A} = \bm{\omega} \times \bm{\xi},
\end{equation}
where $\bm{\omega}$ is the pseudovector whose direction coincides with
$\hat{\bm{\theta}}$, and whose modulus indicates the angular velocity.
This can be easily seen by writing an arbitrary rotation matrix in
exponential form,
\begin{equation}
{\bf R}(t) = e^{\theta(t) {\bf L}^{\hat{\bm{\theta}}} },
\qquad L^{\hat{\bm{\theta}}}_{\alpha \beta} = -\epsilon^{\alpha \beta \gamma} \hat{\theta}_\gamma,
\end{equation}
and by observing that ${\bf A} = {\bf R}^{\rm T} \, \dot{\bf R}(t) \,
\bm{\xi}.$
We have
\begin{equation}
  i \hbar \frac{\partial \cur{\psi}}{\partial t} = \left[ \frac{1}{2 m}
 (\cur{\bf p} - m \bm{\omega} \times \bm{\xi} )^2
  +  V(\bm{\xi}) -
    \frac{1}{2} m ( \bm{\omega} \times \bm{\xi} )^2 \right] \cur{\psi}.
\end{equation}
The Hamiltonian of the system in the rotating frame of reference is,
therefore, identical to that of the system at rest except for two
additional terms: a gauge field and a quadratic potential term. The
latter is unbound from below -- it diverges like $-\rho^2$, where $\rho$
is the distance from the rotation axis.
These two terms have direct classical interpretations as the fictitious
forces (respectively, Coriolis and centrifugal) that appear in the
noninertial rotating frame of reference.
It is interesting to observe that the Coriolis force enters the
Hamiltonian in the exact same way as a uniform magnetic field, with the
only difference that the former acts on the particle mass, while the
latter on its charge.
A magnetic field, in particular, can be described by a gauge field of
the type $$- \frac{q}{2c} {\bf B} \times \bm{\xi}.$$ The above
derivations show that we can obtain the same physical consequences (at
first order in the perturbation amplitude) if, instead of applying a
magnetic field, we rotate the system with an angular velocity equal to
\begin{equation}
\bm{\omega} = \frac{q}{2mc} {\bf B}.
\mylabel{larmor}
\end{equation}
This is, of course, the Larmor frequency.
Thus, in the special case of a rigid rotation, our theory
correctly recovers Larmor's theorem in its known quantum-mechanical
form.~\cite{rmp-gyro}

\subsection{Current density}

The above derivations provide a general picture of how the electronic
Hamiltonian is modified by an arbitrary time-dependent deformation.
Since our main motivation stems from flexoelectricity and, more
generally, from the description of electromechanical phenomena, in this
subsection we shall give special attention to the electronic
current density. This is necessary in order to extract useful information on
the electric polarization
that develops in an insulator following a mechanical deformation.

First of all, we postulate a formula for the current
density that is associated with the Hamiltonian of Eq.~(\ref{schr-xi}),
\begin{equation}
\cur{J}_\beta(\bm{\xi},t) = -\frac{1}{2} g^{\beta \gamma}
\left( -i \cur{\psi}^* \partial_\gamma \cur{\psi} + i \cur{\psi} \partial_\gamma \cur{\psi}^* - 2 A_\gamma | \cur{\psi} |^2 \right).
\label{j-xi}
\end{equation}
Note that $\cur{J}_\beta(\bm{\xi},t)$ describes the
current density \emph{in the curvilinear frame}; this means that the
``convective'' contribution, due to the displacement of the
coordinate frame itself with respect to the laboratory, is
not included. For instance, in the limit of a rigid
rototranslation, the laboratory current ${\bf J}$ is given
by ${\bf J} = {\bf R} \cdot \cur{\bf J} + {\bf v} \rho$,
where ${\bf R}$ is a rotation matrix,
${\bf v} = \dot{\bf r}$ is the velocity and $\rho$ the charge density.

Now, we shall proceed to demonstrate that this formula is indeed
correct.
By ``correct'' we
mean that the probability current satisfies two criteria, namely
(i) the continuity equation,
and (ii) the known transformation laws of the classical four-current in
the nonrelativistic limit.

\subsubsection{Continuity equation}

We need to show that
\begin{equation}
\frac{\partial} {\partial t} \cur{\rho} \Big|_{\bm{\xi}} =
-\bm{\nabla}_{\bm{\xi}} \cdot \cur{\bf J},
\label{eqcont}
\end{equation}
where $\cur{\rho} = -| \cur{\psi} |^2$ is the electronic charge density in the
curvilinear frame. The proof proceeds along the same lines as in the
textbook case of a standard electronic Hamiltonian in the presence of a
vector potential field.
In particular, one needs first to multiply both hand sides of
Eq.~(\ref{schr-xi}) by $\cur{\psi}^*(\bm{\xi},t)$, and then focus on the
real part of the equation by summing each term with its complex
conjugate.
One is left with the time derivative of $\cur{\rho}(\bm{\xi},t)$ on the
left-hand side; after a few manipulations, it is not difficult to show
that the right-hand side corresponds to $-\bm{\nabla}_{\bm{\xi}} \cdot
\cur{\bf J}$.
The only difference with respect to the textbook derivation consists in
the presence of the inverse metric tensor, both in the kinetic energy
operator of Eq.~(\ref{schr-xi}) and in Eq.~(\ref{j-xi}); however, this
does not entail any special complication in the algebra.

As a note of warning, one should keep in mind that the proof of
Eq.~(\ref{eqcont}) is valid only under the key assumption that the
external potential applied to the electrons is \emph{local}.
Thus, the form of the current density as written in Eq.~(\ref{j-xi}) is
inadequate in cases where nonlocal pseudopotentials are adopted in the
calculation.~\cite{cyrus}
This issue, however, is not specific to the present theory of
deformations (it complicates the definition of the current density
already at the level of the standard Cartesian-space Schr\"odinger
equation), and discussing it in detail would lead us far from the scope
of this work.

\subsubsection{Transformation laws}

We next check whether the definition of the current density
that we postulated in Eq.~(\ref{j-xi}) is compatible with the known Galilean
transformation laws of the four-current, which is defined as $J^\mu =
(\rho, J_1, J_2, J_3)$.
In particular, $J^\mu$ transforms as a contravariant vector density,
\begin{equation}
\bar{J}^\mu = \frac{\partial \bar{x}^\mu} {\partial x^\nu} \, J^\nu \,
{\rm det}^{-1}\left[  \frac {\partial\bar{x}^\rho}{\partial x^\sigma} \right],
\end{equation}
where $x^\mu = (t, x_1, x_2, x_3)$ is the coordinate four-vector and the
barred (unbarred) symbols refer to the deformed (original) frame.
In our special case of a nonrelativistic mechanical deformation, we have
$\bar t = t$, and the time is independent of the space coordinates.
By letting the barred and unbarred space coordinates span the Cartesian
${\bf r}$-space and the curvilinear $\bm{\xi}$-space, respectively, we
obtain
\begin{eqnarray}
\rho &=& h^{-1} \cur{\rho} ,  \mylabel{tr-rho} \\
J_l &=&  h^{-1} \left( \cur{\rho} \frac{ \partial {r}_l }{\partial t} + h_{l \beta} \cur{J}_\beta \right).
\mylabel{tr-j}
\end{eqnarray}
The transformation law for the charge density is satisfied by
construction; we need to prove that the same is true for the current
density.

To that end, we write
\begin{equation}
\cur{J}_\beta = -\frac{1}{2}  h^{-1}_ {\beta l} h^{-1}_{\gamma l}
\left( -i \cur{\psi}^* \partial_\gamma \cur{\psi} + i \cur{\psi} \partial_\gamma \cur{\psi}^* -
  2 h_{m \gamma} \frac{\partial r_m}{\partial t} | \cur{\psi} |^2 \right),
\end{equation}
where we have expanded the symbols $g^{\beta \gamma}$ and $B_\gamma$. By
observing that $h^{-1}_{\gamma m} \partial_\gamma = \partial /
\partial r_m$, this can be conveniently rewritten as
\begin{eqnarray}
\cur{J}_\beta &=& -\frac{1}{2}  h^{-1}_ {\beta l}
\left( -i \cur{\psi}^* \frac{\partial \cur{\psi}}{\partial r_l} + i \cur{\psi} \frac{\partial \cur{\psi}^*}{\partial r_l}   -
  2  \frac{\partial r_l}{\partial t} | \cur{\psi} |^2 \right) \nonumber \\
 &=&  h h^{-1}_ {\beta l} \left( J_l - \frac{\partial r_l}{\partial t} \rho \right),
\end{eqnarray}
where
\begin{equation}
J_l = -\frac{1}{2} \left(-i \psi^* \frac{\partial \psi}{\partial r_l}
                                                             +i \psi \frac{\partial \psi^*}{\partial r_l} \right)
\end{equation}
is the probability current in the Cartesian frame.
This is fully consistent with Eq.~(\ref{tr-j}), thus completing the
proof.

\section{Bulk electromechanical response in the linear regime}

In order to make contact with
the linear-response approaches used to describe phenomena
such as piezoelectricity and flexoelectricity, we shall consider,
in the following, a continuous deformation that starts from the
unperturbed state at $t=0$, and occurs slowly enough that it can be
considered small during a finite interval of time following $t=0$.
In such a regime, we can write the elastic deformation as
\begin{equation}
{\bf r} = \bm{\xi} + {\bf u}(\bm{\xi},t),
\mylabel{small-def}
\end{equation}
where both the displacement field, ${\bf u}$, and its time derivative (velocity)
are small.
(This means that, in the linear limit, all terms that are proportional
to $u^2$, $u \dot u$, etc. can be safely dropped.)
We shall also suppose that the deformation is smooth on the scale of the
interatomic spacings. This implies that only the lowest-order gradients
of the displacement field, ${\bf u}(\bm{\xi},t)$, are physically relevant.
Finally, as a reminder, note that we shall only deal with ``clamped-ion''
deformation fields, i.e., we suppose that every atom in the lattice is
displaced by hand according to Eq.~(\ref{clamped}), and neglect any further
relaxation of the individual atomic sublattices.
(Atomic relaxations are, of course, of central importance for a quantitatively
correct description of the electromechanical response.
However, lattice-mediated effects are conceptually simpler
to understand, and have been extensively studied in earlier
publications.
Here we shall only focus on the purely electronic response.)

\subsection{Reciprocal-space analysis}

Without loss of generality, we can represent ${\bf u}(\bm{\xi},t)$ in Fourier space
as a superposition of monochromatic perturbations,
\begin{equation}
{\bf u}(\bm{\xi},t) = \sum_{\bf q \omega} {\bf u}({\bf q},\omega) e^{i{\bf q}\cdot\bm{\xi} - i\omega t}.
\end{equation}
In such a representation, the above conditions on adiabaticity and smoothness can be
formalized by requiring that ${\bf u}({\bf q},\omega)$ appreciably differs from zero
only for small values of $q = |{\bf q}|$ and $\omega$.
In order to derive the electromechanical properties, we shall be concerned with
the electrical current-density as given by Eq.~(\ref{j-xi}), which can be conveniently
represented in Fourier space as well,
\begin{equation}
\cur{\bf J}(\bm{\xi},t) = \sum_{\bf q \omega} \cur{\bf J}({\bf q},\omega) e^{i{\bf q}\cdot\bm{\xi} - i\omega t}.
\mylabel{Fourier-J}
\end{equation}
(For a monochromatic perturbation at a given ${\bf q}$ the microscopic polarization response, ${\bf J}(\bm{\xi},t)$,
generally contains all Fourier components of the type ${\bf G+q}$, where ${\bf G}$ is a
vector of the reciprocal-space Bravais lattice. Here we shall focus on macroscopic effects only,
which are encoded in the ${\bf G=0}$ component.)
Then one can write the relevant coupling coefficients as the linear relationship
between ${\bf u}$ and ${\bf J}$,
\begin{equation}
\cur{\bf J}({\bf q},\omega) = \bm{\chi}^{\bf (J)} ({\bf q},\omega) \cdot {\bf u}({\bf q},\omega),
\mylabel{chi-uJ}
\end{equation}
where $\bm{\chi}^{\bf (J)} ({\bf q},\omega)$ is a $3\times 3$ tensor.

To see how the physical information contained in $\bm{\chi}^{\bf (J)} ({\bf q},\omega)$
relates to the electromechanical (polarization response to a deformation) properties
of the crystal, it is useful to recall the relationship ${\bf J} = -i \omega {\bf P}$,
where the polarization ${\bf P}$ has been Fourier-transformed as in Eq.~(\ref{Fourier-J}).
Then, one can immediately write, for the polarization response in the curvilinear space,
\begin{equation}
\cur{\bf P}({\bf q},\omega) = \bm{\chi}^{\bf (P)} ({\bf q},\omega) \cdot {\bf u}({\bf q},\omega),
\mylabel{chi-uP}
\end{equation}
where we have introduced the electromechanical response function
\begin{equation}
\bm{\chi}^{\bf (P)} = \frac{i}{\omega} \bm{\chi}^{\bf (J)}.
\end{equation}
Finally, one obtains the static clamped-ion electromechanical response as the
adiabatic $\omega \rightarrow 0$ limit of the above,
\begin{equation}
\bm{\chi}^{\bf (P)} ({\bf q}) =
   \bm{\chi}^{\bf (P)} ({\bf q},\omega=0) = i \frac{\partial \bm{\chi}^{\bf (J)} ({\bf q},\omega) }{\partial \omega} \Big|_{\omega=0}.
\mylabel{chi-PJ}
\end{equation}
This procedure reflects the fundamental physical nature of the
electrical polarization, which is understood as the time integral of the
transient current density that flows through the sample in the course of
an adiabatic transformation of the crystal.

Note that $\bm{\chi}^{\bf (P)} ({\bf q})$ has a direct relationship to the polarization
response tensors that were considered in earlier works, e.g.,
\begin{equation}
\chi_{\alpha\beta}^{\bf (P)} ({\bf q}) = \sum_\kappa \overline{P}_{\alpha,\kappa\beta}^{\bf q},
\end{equation}
where $\overline{P}_{\alpha,\kappa\beta}^{\bf q}$ describes the contribution of a modulated
displacement (along $\beta$) of the atomic sublattice $\kappa$ to the macroscopic polarization
along $\alpha$ [see Eq.~(13) of Ref.~\onlinecite{chapter-15}].
(The nuclear point charges are included in
$\overline{P}_{\alpha,\kappa\beta}^{\bf q}$, following the
original definition,\cite{artlin,chapter-15} while they are
absent from $\chi_{\alpha\beta}^{\bf (P)}$ by construction -- in
the curvilinear space, the atoms do not move from their original
location, and hence do not produce any current therein.)

\subsection{Perturbation theory}

To calculate $\bm{\chi}^{\bf (J)}$ in a quantum-mechanical context,
we shall first derive (in real space) the
current density response to a monochromatic perturbation of the type
${\bf u}({\bf r},t) = \bm{\lambda} e^{i{\bf q}\cdot{\bf r} - i\omega t}$,
and subsequently select its lowest Fourier component, as required by the
present macroscopic context.
Even if the following derivations will be carried out in curvilinear space,
as there is no longer a potential risk of confusion we shall indicate the
real-space coordinate as ${\bf r}$ and omit the ``$\cur{\,\,\,}$''
symbol  henceforth.

Consider the unperturbed single-particle density operator,
\begin{equation}
\hat{\mathcal{P}}^{(0)} = \sum_n | \psi^{(0)}_n \rangle f^{(0)}_n \langle \psi^{(0)}_n |,
\end{equation}
where $\psi^{(0)}_n({\bf r}) $
are eigenstates of the unperturbed Hamiltonian ,
\begin{equation}
\hat{\mathcal{H}}^{(0)} | \psi^{(0)}_n \rangle = \epsilon^{(0)}_n | \psi^{(0)}_n \rangle.
\end{equation}
($f^{(0)}_n$ indicates the occupation of the state, which is either 0 or 1
for an insulating crystal in its electronic ground state.)
In presence of the perturbation, the dynamical evolution of the density matrix is
described by the single-particle Liouville equation
\begin{equation}
i \hbar \frac{ \partial \hat{\mathcal{P}} }{\partial t} = [\hat{\mathcal{H}}(t), \hat{\mathcal{P}}],
\mylabel{liouville}
\end{equation}
where $\hat{\mathcal{H}}(t)$ is the curvilinear-frame Hamiltonian of Eq.~(\ref{schr-xi}).
(Earlier derivations of the first-order adiabatic current based on
the single-particle density matrix can be found in Refs.~\onlinecite{adler}
and~\onlinecite{niu/thouless-84}.)
Then, we can rewrite the current density of Eq.~(\ref{j-xi}) as
\begin{equation}
J_\alpha ({\bf r},t) = g^{\alpha \beta} \, {\rm Tr}  (  \hat{\mathcal{J}}_\beta \hat{\mathcal{P}} ),
\mylabel{j-new}
\end{equation}
where $g^{\alpha \beta}$, as usual, refers to the inverse metric tensor
(implicit summation over $\beta$ is assumed), the sum runs over the valence wavefunctions, and
we have introduced the ``curvilinear'' current-density operator $\hat{\mathcal{J}}_\alpha$,
%
\begin{equation}
 \hat{\mathcal{J}}_\beta ({\bf r},t)= -\frac{\hat{p}_\beta |{\bf r} \rangle \langle {\bf r} | + |{\bf r} \rangle \langle {\bf r} | \hat{p}_\beta}{2}
  +  |{\bf r} \rangle A_\alpha ({\bf r},t) \langle {\bf r} | .
\mylabel{j-op}
\end{equation}
The minus sign appears, as in Eq.~(\ref{j-xi}), because 
in our units the charge of the electron is $-1$.
Note that $\hat{\mathcal{J}}_\alpha$
explicitly depends on space and time
via the effective gauge potential ${\bf A}({\bf r},t)$
of Eq.~(\ref{Adef}). Time dependence is
also implicitly present in $\hat{\mathcal{P}}$ via Eq.~(\ref{liouville}).

%

%
%
In the linear regime, we can expand both operators, $\mathcal{J}$ and
$\mathcal{P}$, in powers of the displacement amplitude, $\bm{\lambda}$,
\begin{eqnarray}
\hat{\mathcal{J}}_\alpha  &=& \hat{\mathcal{J}}_\alpha^{(0)} +  \lambda_\beta \hat{\mathcal{J}}_\alpha^{(\lambda_\beta)} + \cdots, \\
\hat{\mathcal{P}} &=& \hat{\mathcal{P}}^{(0)} + \lambda_\beta   \hat{\mathcal{P}}^{(\lambda_\beta)} + \cdots,
\end{eqnarray}
where the dots stand for higher-order terms that have been dropped.
By incorporating the above expansions into Eq.~(\ref{j-new}) we readily obtain
\begin{equation}
\frac{\partial J_{\alpha} ({\bf r},t)}{\partial \lambda_\beta} = {\rm Tr}  \left(  \hat{\mathcal{J}}_\alpha^{(\lambda_\beta)} \hat{\mathcal{P}}^{(0)} \right) +
   {\rm Tr}  \left(  \hat{\mathcal{J}}_\alpha^{(0)} \hat{\mathcal{P}}^{(\lambda_\beta)} \right).
  \mylabel{j-new2}
\end{equation}
[Note that the inverse metric tensor of Eq.~(\ref{j-new}) also depends on $\lambda_\beta$,
which in principle would generate an extra term; however, one can easily see that the
first-order expansion of $g^{\alpha \beta}$ does not contribute to the current density in
a time-reversal symmetric crystal -- there are no circulating currents in the ground state.
Thus, in the present context $g^{\alpha \beta}$ can be safely replaced with a Kronecker
delta.]

The expansion of the current-density operator of Eq.~(\ref{j-op})
is relatively straightforward after observing that, in the linear limit,
Eq.~(\ref{Adef}) gives
${\bf A} = \dot{\bf u} = -i\omega \bm{\lambda} e^{i{\bf q}\cdot{\bf r} - i\omega t}$;
we obtain
\begin{eqnarray}
\hat{\mathcal{J}}^{(0)}_\alpha ({\bf r}) &=& -\frac{\hat{p}_\alpha |{\bf r} \rangle \langle {\bf r} | + |{\bf r} \rangle \langle {\bf r} | \hat{p}_\alpha}{2}, \\
\hat{\mathcal{J}}_\alpha^{(\lambda_\beta)}({\bf r},t) &=&
  -i\omega \delta_{\alpha \beta} \, |{\bf r} \rangle e^{i{\bf q}\cdot{\bf r} - i\omega t} \langle {\bf r} | .
\end{eqnarray}
At order zero, we correctly recover the standard textbook expression for the
current-density operator in a Cartesian space, which does not depend explicitly on time,
while at first order we have a real-space projection operator times some complex prefactors.
The only remaining task is now to derive an explicit formula for the first-order density matrix,
$\hat{\mathcal{P}}^{(\lambda_\beta)}$, which we shall do in the following paragraphs.

By linearizing the Liouville equation, Eq.~(\ref{liouville}), and by assuming that the
time dependence of the response is the same as that of the perturbing field, we
easily arrive at
\begin{equation}
\langle \psi_m | \hat{\mathcal{P}}^{(\lambda_\beta)} | \psi_n \rangle =
  \frac{ \langle \psi_m | \hat{\mathcal{H}}^{(\lambda_\beta)} | \psi_n \rangle (f_n - f_m)}
       {  \epsilon_n - \epsilon_m + \omega  },
\end{equation}
where we have dropped the superscript ``$(0)$'' on the wavefunctions, eigenvalues
and occupancies to simplify the notation, and $\hat{\mathcal{H}}^{(\lambda_\beta)}$
relates to the expansion of the Hamiltonian operator in powers of $\bm{\lambda}$,
\begin{equation}
\hat{\mathcal{H}} = \hat{\mathcal{H}}^{(0)} + \lambda_\beta   \hat{\mathcal{H}}^{(\lambda_\beta)} + \cdots.
\end{equation}
(An explicit expression of $\hat{\mathcal{H}}^{(\lambda_\beta)}$ is derived in the Appendix.)
We thus arrive at a closed expression for the
current density of Eq.~(\ref{j-new2}),
\begin{eqnarray}
\frac{\partial J_{\alpha} ({\bf r},t)}{\partial \lambda_\beta} &&= i\omega \delta_{\alpha \beta}
                     e^{i{\bf q}\cdot{\bf r} - i\omega t} \rho^{(0)}({\bf r}) + \mylabel{j-new3} \\
 && \sum_{mn} \frac{ \langle \psi_n |  \hat{\mathcal{J}}^{(0)}_\alpha ({\bf r}) | \psi_m \rangle
     \langle \psi_m | \hat{\mathcal{H}}^{(\lambda_\beta)} | \psi_n \rangle (f_n - f_m)}
       {  \epsilon_n - \epsilon_m + \omega  }, \nonumber
\end{eqnarray}
where $\rho^{(0)}({\bf r})= -\sum_n |\psi_n({\bf r})|^2$ is the
ground-state electronic charge density.

We shall now filter out the macroscopic component of the response
at wavevector ${\bf q}$ and
eliminate the trivial phase factor of $e^{- i\omega t}$.
After performing both operations, we obtain the desired current-response function as
a Fourier transform
\begin{equation}
\chi_{\alpha \beta}^{\bf (J)} ({\bf q},\omega) =  \frac{e^{i\omega t}}{\Omega} \int_{\rm cell} d^3r e^{-i{\bf q}\cdot{\bf r}}
\frac{\partial J_{\alpha} ({\bf r},t)}{\partial \lambda_\beta}.
\mylabel{fourier-mac}
\end{equation}
By combining Eq.~(\ref{fourier-mac}) with Eq.~(\ref{j-new3}), we arrive at
\begin{eqnarray}
\chi_{\alpha \beta}^{\bf (J)} && ({\bf q},\omega)
 = -i \omega \frac{N}{\Omega} \delta_{\alpha \beta} + e^{i\omega t} \times  \mylabel{j-new4} \\
    && \sum_{mn} \frac{ \langle \psi_n |  \hat{\mathcal{J}}^{(0)}_\alpha ({\bf q}) | \psi_m \rangle
     \langle \psi_m | \hat{\mathcal{H}}^{(\lambda_\beta)} | \psi_n \rangle (f_n - f_m)}
       {  \epsilon_n - \epsilon_m + \omega  },\nonumber
\end{eqnarray}
where we have introduced the reciprocal-space representation of the unperturbed
current-density operator
\begin{eqnarray}
\hat{\mathcal{J}}_\alpha^{(0)} ({\bf q}) &=& - \frac{1}{\Omega} \int d^3r
     \frac{ e^{-i {\bf q} \cdot {\bf r}}  \hat{p}_\alpha + \hat{p}_\alpha e^{-i {\bf q} \cdot {\bf r}}  }{2}
                                             \nonumber\\
                               &=& - \frac{1}{\Omega} \int d^3r  \left( \hat{p}_\alpha + \frac{q_\alpha}{2} \right) \, e^{-i {\bf q} \cdot {\bf r}},
                               \mylabel{j-0}
\end{eqnarray}
and $N$ is the number of valence electrons in the primitive cell.

As a last step, it is useful to bring Eq.~(\ref{j-new4}) into a simpler
form by observing that the occupation factor $(f_n - f_m)$ only
selects cross-gap matrix elements.
Thus, we can introduce the first-order wavefunction response
to the perturbation at the frequency $\omega$ as a sum over
conduction states ($c$) only,
\begin{equation}
| \psi_v^{(\lambda_\beta)} ({\bf q},\omega) \rangle
     =  e^{i\omega t} \sum_{c}  | \psi_c \rangle \frac{ \langle \psi_c | \hat{\mathcal{H}}^{(\lambda_\beta)}({\bf q},\omega) | \psi_v \rangle}
  {\epsilon_v - \epsilon_c + \omega},
  \mylabel{adiab}
\end{equation}
where we have made the $({\bf q},\omega)$-dependence of the first-order Hamiltonian explicit,
and rewrite Eq.~(\ref{j-new4}) as a summation over valence states ($v$),
\begin{eqnarray}
\chi_{\alpha \beta}^{\bf (J)}  ({\bf q},\omega)
 &=& -i \omega \frac{N}{\Omega} \delta_{\alpha \beta} + \nonumber \\
    && \sum_{v} \Big\{ \langle \psi_v |  \hat{\mathcal{J}}^{(0)}_\alpha ({\bf q}) |  \psi_v^{(\lambda_\beta)} ({\bf q},\omega) \rangle + \mylabel{j-new5} \\
    &&  \,\,\,\,  \langle \psi_v^{(\lambda_\beta)} ({\bf -q},-\omega) |  \left[\hat{\mathcal{J}}^{(0)}_\alpha (-{\bf q}) \right]^\dagger |  \psi_v  \rangle
    \Big\}. \nonumber
\end{eqnarray}
To arrive from Eq.~(\ref{j-new4}) to Eq.~(\ref{j-new5}) we have used the following general
property of any first-order Hamiltonian that is associated with a monochromatic perturbation,
\begin{equation}
\left[\hat{\mathcal{H}}^{(1)} ({\bf q},\omega) \right]^\dagger = \hat{\mathcal{H}}^{(1)} (-{\bf q},-\omega) \mylabel{H1-dag}
\end{equation}
which follows from the continued Hermiticity of
$\hat{\mathcal{H}}({\bf r},t)$ in the presence of the perturbation.
Note that the current operator is related to the first-order Hamiltonian
in presence of an electromagnetic vector potential field, ${\bf A}$,
\begin{equation}
\hat{\mathcal{J}}^{(0)}_\alpha ({\bf q}) = - \left[ \hat{\mathcal{H}}^{(A_\alpha)} ({\bf q}) \right]^\dagger.
\mylabel{J-HA}
\end{equation}
This means that Eq.~(\ref{H1-dag}) holds for the current operator as well, thus
completing the proof of Eq.~(\ref{j-new5}).

In summary, the heavy algebra of this section has provided us with an important
result for the current response function, $\chi_{\alpha \beta}^{\bf (J)}$. To
clarify what we have achieved so far, it is useful to rewrite Eq.~(\ref{j-new5}) as
\begin{equation}
\chi_{\alpha \beta}^{\bf (J)}  ({\bf q},\omega)
 = -i \omega \frac{N}{\Omega} \delta_{\alpha \beta} + F_{\alpha \beta} ({\bf q},\omega) + F^*_{\alpha \beta} (-{\bf q},-\omega),
  \mylabel{chi-F}
\end{equation}
where the auxiliary functions $F_{\alpha \beta}$ can be expressed as sums over
occupied-state indices only,
\begin{equation}
F_{\alpha \beta} ({\bf q},\omega) =
   \sum_{v}  \langle \psi_v |  \hat{\mathcal{J}}^{(0)}_\alpha ({\bf q}) |  \psi_v^{(\lambda_\beta)} ({\bf q},\omega) \rangle.
\mylabel{fqw}
\end{equation}

At first sight, our progress towards a numerically tractable
theory might appear only cosmetic, since an infinite sum over unoccupied
states is still present in the definition of
the first-order wavefunctions in Eq.~(\ref{adiab}).
This is, in principle, undesirable
from the point of view of an implementation.
%
However, expressions like Eq.~(\ref{adiab}) can easily be replaced, in the
context of density-functional perturbation theory, by computationally more
palatable Sternheimer equations.


%
%

\subsection{Frequency expansion}

We shall now extract the static polarization response function, $\chi_{\alpha \beta}^{\bf (P)} ({\bf q})$,
by taking the zero-frequency limit of the current response according to
Eq.~(\ref{chi-PJ}).  Substituting Eq.~(\ref{chi-F})
and taking note of the fact that
$F_{\alpha \beta} ({\bf q},\omega)=F^*_{\alpha \beta} (-{\bf q},-\omega)$,
which follows from the assumption of time-reversal symmetry,
we obtain
\begin{equation}
\chi_{\alpha\beta}^{\bf (P)} ({\bf q}) = \frac{N}{\Omega}\delta_{\alpha \beta}
  +2i \frac{\partial F_{\alpha\beta}({\bf q},\omega) }{\partial \omega} \Big|_{\omega=0}.
\mylabel{chiPint}
\end{equation}

Our next task, then, is
to work out the frequency expansion of the auxiliary function
${F}_{\alpha \beta} ({\bf q},\omega)$, which in turn
depends on $\omega$ via the first-order wavefunctions of Eq.~(\ref{adiab}).
%
We shall, first of all, separate the ``static'' (frequency-independent) and
``dynamic'' contributions to the first-order Hamiltonian,
\begin{equation}
e^{i\omega t} \hat{\mathcal{H}}^{(\lambda_\beta)}({\bf q}, \omega) = \hat{\mathcal{H}}^{(\lambda_\beta)}({\bf q})
 - i\omega \hat{\mathcal{H}}^{(\dot{\lambda}_\beta)}({\bf q}),
\mylabel{wexp-H1}
\end{equation}
where we have set $\omega=0$ in the first term, and collected the remainder
in the second term.  (Note that there are no other terms, e.g.,
dependent on $\omega^2$, as we are working within the linear
approximation in the displacement field amplitude.)
By combining Eq.~(\ref{wexp-H1}) and Eq.~(\ref{adiab}) we obtain then,
for the wavefunction response,
\begin{eqnarray}
|\psi_{v}^{(\lambda_\beta)}({\bf q},\omega)  \rangle &=&
|\psi_{v}^{(\lambda_\beta)}({\bf q},\omega = 0) \rangle
    -i\omega  | \delta \psi_{v}^{(\lambda_\beta)}({\bf q}) \rangle \nonumber \\
 &&   - i\omega  | \psi_{v}^{(\dot{\lambda}_\beta)}({\bf q}) \rangle + \cdots,
\mylabel{3terms}
\end{eqnarray}
where the second and third terms originate, respectively, from the frequency
expansion
of the energy denominator in Eq.~(\ref{adiab}),
\begin{eqnarray}
 | \delta \psi_{v}^{(\lambda_\beta)} ({\bf q}) \rangle
     = -i \sum_{c}  | \psi_{c} \rangle \frac{ \langle \psi_{c} | \hat{\mathcal{H}}^{(\lambda_\beta)}({\bf q}) | \psi_{v} \rangle}
  {(\epsilon_{v} - \epsilon_{c})^2}. \mylabel{delpsi}
 \end{eqnarray}
 and of the first-order Hamiltonian, Eq.~(\ref{wexp-H1}),
\begin{equation}
| \psi_{v}^{(\dot{\lambda}_\beta)}({\bf q}) \rangle =
     \sum_{c}  | \psi_{c} \rangle \frac{ \langle \psi_{c} | \hat{\mathcal{H}}^{(\dot{\lambda}_\beta)}({\bf q}) | \psi_{v} \rangle}
  {\epsilon_{v} - \epsilon_{c}}. \mylabel{psi1}
\end{equation}

Note that Eq.~(\ref{psi1}) is very similar in form to Eq.~(\ref{delpsi}), except for the
power of two in the denominator and the factor of $-i$ appearing in the latter.
In fact, one can show that $| \delta \psi_{v}^{(\lambda_\beta)}({\bf q})  \rangle $ is directly
related to the \emph{adiabatic} wavefunction response, at first order in the velocity,
to the ``static'' perturbation $\hat{\mathcal{H}}^{(\lambda_\beta)}({\bf q})$, when such a perturbation
is slowly switched on as a function of time.
To see this, one can go back to the Liouville equation, Eq.~(\ref{liouville}), and
perform an expansion in the velocity of the perturbation, rather than its amplitude.
First we write
\begin{equation}
i \hbar \dot{\lambda} \frac{ \partial \hat{\mathcal{P}} (\lambda)}{\partial \lambda} = [\hat{\mathcal{H}}(\lambda), \hat{\mathcal{P}}(\lambda)],
\end{equation}
and use a trial solution of the type
\begin{equation}
\hat{\mathcal{P}} (\lambda) \simeq \hat{\mathcal{P}}^{(0)} (\lambda) + \dot{\lambda} \hat{\mathcal{P}}^{(1)} (\lambda).
\end{equation}
Next, by expanding in powers of $\dot{\lambda}$ we have, at order zero, the usual adiabatic
limit of the quantum system following its instantaneous ground state,
\begin{equation}
[\hat{\mathcal{H}}(\lambda), \hat{\mathcal{P}}^{(0)}(\lambda)] = 0.
\end{equation}
Finally, at first order in $\dot{\lambda}$, we obtain
\begin{equation}
i \partial_\lambda \hat{\mathcal{P}}^{(0)}(\lambda) = [\hat{\mathcal{H}}(\lambda), \hat{\mathcal{P}}^{(1)}(\lambda)],
\end{equation}
which after projecting over a basis of instantaneous eigenstates of $\hat{\mathcal{H}}(\lambda)$ leads to
\begin{equation}
\langle \psi_m | \hat{\mathcal{P}}^{(1)} |\psi_n \rangle = -i \frac{ \langle \psi_m | \partial_\lambda \hat{ \mathcal{P} }^{(0)} |\psi_n \rangle }
    {\epsilon_n - \epsilon_m}.
\end{equation}
(We have omitted the obvious parametric dependence on $\lambda$ of all quantities in the above equation.)
This result illustrates the physical meaning of the additional energy denominator and the factor of $-i$
in Eq.~(\ref{delpsi}).

Returning to our main argument,
we are ready to carry out the expansion of ${F}_{\alpha \beta} ({\bf q},\omega)$.
Plugging Eq.~(\ref{3terms}) into Eq.~(\ref{fqw}),
we obtain
\begin{equation}
  F_{\alpha \beta} ({\bf q},\omega) = f_{\alpha \beta}({\bf q})
  - i\omega \left[ \bar{g}_{\alpha \beta}({\bf q}) + \Delta {g}_{\alpha \beta}({\bf q}) \right] + \cdots,
 \mylabel{wexp-f}
\end{equation}
where the three contributions derive from the three terms on the
right-hand side of Eq.~(\ref{3terms}) respectively.
That is,
$f_{\alpha \beta}({\bf q}) = F_{\alpha \beta} ({\bf q},0)$ and
\begin{eqnarray}
\bar{g}_{\alpha \beta}({\bf q}) &=&  \sum_{v}   \langle \psi_v |  \hat{\mathcal{J}}^{(0)}_\alpha({\bf q})  | \delta \psi_v^{(\lambda_\beta)}({\bf q}) \rangle,
\mylabel{gbarq} \\
\Delta {g}_{\alpha \beta}({\bf q}) &=&  \sum_{v}   \langle \psi_v |  \hat{\mathcal{J}}^{(0)}_\alpha({\bf q})  | \psi_v^{(\dot{\lambda}_\beta)}({\bf q}) \rangle. \mylabel{gq}
\end{eqnarray}
We shall refer to the above responses as as \emph{static}
and \emph{dynamic} respectively.
   It is important to note, in this context, that the words 
   ``static'' and ``dynamic'' do not refer 
   to the physical nature of these terms: Indeed, both
   contribute to the bulk polarization field that results from
   a \emph{static} strain gradient; also, both pieces contribute
   to the (transient) macroscopic current that flows at the bulk
   level when the deformation is applied \emph{dynamically}.
   Instead, this nomenclature is motivated by the 
   mathematical origin of these contributions, which stem 
   respectively from the first-order variation of the Hamiltonian 
   operator, and from the effect of the coordinate transformation on
   the time derivative (see Appendix~\ref{app:a}).

After plugging Eq.~(\ref{wexp-f}) into
Eq.~(\ref{chiPint}),
we finally obtain the polarization response function as
\begin{equation}
\chi_{\alpha \beta}^{\bf (P)}({\bf q}) = \bar{\chi}_{\alpha \beta}({\bf q}) + \Delta \chi_{\alpha \beta}({\bf q}).
\mylabel{chitot}
\end{equation}
Here $\bar{\chi}_{\alpha \beta}$ is the static part encoding the contribution of the
static first-order Hamiltonian via $| \delta \psi_v^{(u_\beta)}({\bf q}) \rangle$,
\begin{equation}
\bar{\chi}_{\alpha \beta}({\bf q}) =  2 \bar{g}_{\alpha \beta}({\bf q}), \mylabel{stat-chi}
\end{equation}
while the remainder in Eq.~(\ref{chitot}) is the dynamic part,
\begin{equation}
\Delta \chi_{\alpha \beta}({\bf q}) = 2 \Delta {g}_{\alpha \beta}({\bf q})
                                     + \frac{N}{\Omega}\delta_{\alpha \beta}, \mylabel{dyn-chi}
\end{equation}
which arises due to the effective vector-potential field that appears in the
time-dependent Schr\"odinger equation, Eq.~(\ref{schr-xi}), as a result of the coordinate
transformation. 
Indeed, as we shall see shortly, the perturbing operator $\hat{\mathcal{H}}^{(\dot{\lambda}_\beta)}({\bf q})$ of Eqs.~(\ref{wexp-H1}) and~(\ref{psi1}) corresponds to \emph{minus} the first-order Hamiltonian in a vector potential field, 
\begin{equation}
\hat{\mathcal{H}}^{(\dot{\lambda}_\beta)}({\bf q}) = -\hat{\mathcal{H}}^{(A_\beta)}({\bf q}).
\mylabel{lb-Ab}
\end{equation}
(The minus sign comes from the negative electron charge, which implies that the velocity operator is $\hat{v}_\beta = \hat{p}_\beta + A_\beta$ in the electromagnetic case.) For this reason, we shall refer to this contribution as either ``gauge-field'', ``vector-potential'' or ``dynamic'' henceforth.

The dynamic contribution
$\Delta \chi_{\alpha \beta}({\bf q})$ is unusual in
the context of the existing literature, and deserves further attention.
The clear priority at this point is to understand whether it
produces any contribution to the macroscopic electromechanical
tensors, and whether such contribution can be related somehow to
some well-defined (and possibly measurable) property of the material.
We shall primarily focus on this task in the remainder of the manuscript.

\subsection{Long-wave expansion}

After dealing with the linear expansion in the deformation amplitude
and frequency (above subsections), there is one last step that we need
to take care of in order to arrive at the macroscopic electromechanical
tensors -- the long-wave expansion of $\bm{\chi}^{\bf (P)} ({\bf q})$ in
powers of the wavevector ${\bf q}$.
This readily yields the clamped-ion piezoelectric (${\bf e}$) and flexoelectric ($\bm{\mu}$)
tensors at first and second order in ${\bf q}$ respectively,
\begin{equation}
\chi_{\alpha \beta}^{\bf (P)}({\bf q}) = i q_\gamma e_{\alpha, \beta \gamma} -
 q_\gamma q_\delta \mu_{\alpha \beta, \gamma \delta} + \cdots.
 \mylabel{chi-expq}
 \end{equation}
We shall separately discuss the expansion of $\bar{\chi}$ and $\Delta \chi$
in the following, highlighting their respective contribution to the
aforementioned tensors.

Before doing so, we need to remove the incommensurate
phases from the operators and wavefunctions, as they are problematic in
the context of a parametric ${\bf q}$-expansion; the standard approach to
deal with this issue is to introduce a crystal momentum representation.
%
%
%
For the ground-state orbitals we have
\begin{eqnarray}
|\psi_{n\bf k}\rangle &=& e^{i {\bf k\cdot r}} | \phi_{n\bf k} \rangle,
\end{eqnarray}
where $\phi_{n\bf k}$ are cell-periodic functions.
Then, all the sums over occupied states of the previous sections
need to be replaced by
a sum over valence bands plus a Brillouin-zone average,
\begin{equation}
\sum_v \rightarrow \sum_n \int [ d^3 k],
\end{equation}
where we have introduced the short-hand notation $[ d^3 k] = \Omega /(2\pi)^3 d^3 k$.
Note that the first-order wavefunctions
contain a shift in momentum space by ${\bf q}$, which reflects the monochromatic
nature of the perturbation,
\begin{equation}
| \psi_{n \bf k}^{(1)} ({\bf q}) \rangle = e^{i {\bf {(k+q)}\cdot r}} | \phi_{n\bf k,q}^{(1)} \rangle,
\end{equation}
Finally, the cell-periodic operators are constructed in order to
conveniently reabsorb the above phase factors,
\begin{eqnarray}
\hat{\mathcal{H}}_{\bf k,q}^{(1)}  &=&  e^{-i {\bf (k+q) \cdot r} } \, \hat{\mathcal{H}}^{(1)}({\bf q})  \, e^{i {\bf k\cdot r}}.
\end{eqnarray}
Note that, consistent with Eq.~(\ref{J-HA}), we shall define
\begin{equation}
\hat{\mathcal{J}}^{(0)}_{\alpha \bf k,q} = -\left(\hat{\mathcal{H}}_{\bf k,q}^{(A_\alpha)} \right)^\dagger =
   e^{-i {\bf k\cdot r}} \hat{\mathcal{J}}^{(0)}_{\alpha} ({\bf q}) e^{i {\bf (k+q) \cdot r} }.
   \mylabel{J-HAk}
\end{equation}
Equation~(\ref{H1-dag}) becomes
\begin{equation}
(\hat{\mathcal{O}}_{\bf k,q})^\dagger =  \hat{\mathcal{O}}_{\bf k+q,-q}
\end{equation}
where $\hat{\mathcal{O}}$ stands for either $\hat{\mathcal{J}}^{(0)}_{\alpha}$ or a generic
first-order Hamiltonian $\hat{\mathcal{H}}^{(1)}$.

\subsubsection{Static contribution}

\mylabel{sec:staticpert}

Regarding the static part $\bar{\chi}$, we defer the detailed derivation of
the operator $\hat{\mathcal{H}}_{\bf k,q}^{(\lambda_\beta)}$ to Appendix~\ref{app:b}.
Here we shall limit ourselves to using some key
properties of its small-${\bf q}$ expansion,
\begin{equation}
\hat{\mathcal{H}}_{\bf k,q}^{(\lambda_\beta)} = i q_\gamma \hat{\mathcal{H}}_{\bf k}^{(\beta \gamma)}
   - q_\gamma q_\delta \hat{\mathcal{H}}_{\bf k}^{(\beta, \gamma \delta)} + \cdots,
\end{equation}
which we summarize as follows:
\begin{itemize}
\item $\hat{\mathcal{H}}_{\bf k,q}^{(\lambda_\beta)}$ vanishes at ${\bf q}=0$. This has to do with the fact that the ${\bf q}\rightarrow 0$ limit
of a monochromatic displacement wave is a rigid translation, and a rigid translation has no effect whatsoever on
the static physical properties of the crystal.
\item The first-order term $\hat{\mathcal{H}}_{\bf k}^{(\beta \gamma)}$ is symmetric with respect to $\beta \gamma$
exchange, and corresponds to the \emph{uniform strain} perturbation of Ref.~\onlinecite{hamann-metric}.
\item Both properties propagate to the first-order static and adiabatic wavefunctions, which can be expanded as
\begin{eqnarray}
 | \phi_{n\bf k}^{(\lambda_\beta)}({\bf q}) \rangle &=& i q_\gamma | \phi_{n\bf k}^{(\beta\gamma)} \rangle + \cdots, \\
 | \delta \phi_{n\bf k}^{(\lambda_\beta)}({\bf q}) \rangle &=& i q_\gamma |\delta \phi_{n\bf k}^{(\beta\gamma)} \rangle + \cdots. \mylabel{oq1-adiab}
\end{eqnarray}
The functions $| \phi_{n\bf k}^{(\beta\gamma)} \rangle$, in particular, correspond to
the strain response functions $| \psi_{n\bf k}^{(\eta_{\beta\gamma})} \rangle$ of Ref.~\onlinecite{hamann-metric}.
\end{itemize}
The above considerations readily yield, 
by combining Eqs.~(\ref{gbarq}), (\ref{stat-chi}),
 (\ref{chi-expq}) and (\ref{oq1-adiab}),
an explicit formula for the contribution
of $\bar{\chi}$ to the piezoelectric tensor,
\begin{equation}
\bar{e}_{\alpha,\beta \gamma} =
    2  \int [ d^3 k] \sum_{v} \langle \phi_{v \bf k} |  \hat{\mathcal{J}}^{(0)}_{\bf k \alpha}  | \delta \phi_{v \bf k}^{(\beta\gamma)} \rangle,
\end{equation}
where $\hat{\mathcal{J}}^{(0)}_{\bf k \alpha}  =- \partial \hat{\mathcal{H}}_{\bf k}^{(0)} / \partial k_\alpha$
is the macroscopic current operator. This is easily
shown to match Eq.~(16) of Ref.~\onlinecite{hamann-metric}
by rearranging the energy denominators,
\begin{eqnarray}
 \sum_{v} && \langle \phi_{v \bf k} |  \hat{\mathcal{J}}^{(0)}_{\bf k \alpha}  | \delta \phi_{v \bf k}^{(\beta\gamma)} \rangle \nonumber \\
 && =  -i \sum_{vc} \frac { \langle  \phi_{v \bf k} |  \hat{\mathcal{J}}^{(0)}_{\bf k \alpha}     | \phi_{c \bf k} \rangle
                             \langle  \phi_{c \bf k} |  \hat{\mathcal{H}}_{\bf k}^{(\beta \gamma)} | \phi_{v \bf k} \rangle }
                           { ( \epsilon_{v\bf k} - \epsilon_{c \bf k})^2} \nonumber \\
 && = - \sum_{v} \langle i \tilde{\partial}_\alpha  \phi_{v \bf k} |  \phi_{v \bf k}^{(\beta\gamma)} \rangle,
\end{eqnarray}
where we have introduced the standard definition of the auxiliary ``$d/dk$'' wavefunctions,
\begin{equation}
|i \tilde{\partial}_\alpha  \phi_{v \bf k} \rangle = i \sum_c | \phi_{c \bf k} \rangle
    \frac{  \langle  \phi_{c \bf k} |  \partial \hat{\mathcal{H}}_{\bf k}^{(0)} / \partial k_\alpha | \phi_{v \bf k} \rangle }
         { \epsilon_{v\bf k} - \epsilon_{c \bf k} }.
\end{equation}
Thus, the present theory yields the widely accepted formula for the clamped-ion
piezoelectric response as a long-wave expansion of the static contribution to
the electromechanical response.

By pushing the ${\bf q}$-expansion to second order [recall Eq.~(\ref{chi-expq})], 
one can readily access the static~\footnote{We stress that the ``static'' and
  ``dynamic'' attributes that we use in this work in the context of the purely 
  electronic response have nothing to do with the mass dependence of the 
  lattice-mediated contribution, which was discussed in earlier works.~\cite{artlin}}
contribution to the clamped-ion flexoelectric tensor, $\bar{\bm{\mu}}$.
While the resulting formulas can be derived analytically, they
are significantly more complex (e.g., both the contribution of the
uniform strain and strain gradient response functions need, in 
principle, to be taken into account), and their 
physical interpretation is not as obvious as in the piezoelectric case. 
From the point of view of the code implementation it might
be convenient to calculate, instead, the electromechanical response at
finite ${\bf q}$, and later take the long-wave expansion of Eq.~(\ref{chi-expq})
numerically; we took such an approach in Ref.~\onlinecite{cyrus}.

\subsubsection{Dynamic contribution}

We now elaborate on the dynamic term and derive its
contributions to the piezoelectric and flexoelectric tensor.
First of all, we need an explicit expression for the operators that
are implicitly involved in Eq.~(\ref{dyn-chi}).
By combining Eq.~(\ref{lb-Ab}) with Eq.~(\ref{J-HAk}) we readily obtain
\begin{equation}
\hat{\mathcal{J}}_{\alpha\bf k,q } = \hat{\mathcal{H}}_{\bf k,q}^{(\dot{\lambda}_\alpha)} = - \left( \hat{p}_{\bf k \alpha} +\frac{q_\alpha}{2} \right),
\end{equation}
\st{Then,} 
We can then write a closed expression for the intermediate function $\Delta g_{\alpha \beta}({\bf q})$,
\begin{widetext}
\begin{equation}
 \Delta g_{\alpha \beta}({\bf q}) = -\int [ d^3 k]  \, \sum_{nc} \frac{ 
  \langle u_{n\bf k} | \left( \hat{p}_{{\bf k} \alpha} + q_\alpha/2 \right) |u_{c{\bf k+q}} \rangle
  \langle u_{c{\bf k+q}} | \left( \hat{p}_{{\bf k}\beta} + q_\beta/2 \right) |u_{n{\bf k}} \rangle
 }{\epsilon_{c{\bf k+q}} - \epsilon_{n{\bf k}} },
     \mylabel{gkq}
\end{equation}
\end{widetext}
which is clearly Hermitian in the Cartesian indices.
($n$ and $c$ run, as usual, over valence and conduction states, respectively).
Note that $\Delta \chi_{\alpha \beta}({\bf q})$ of Eq.~(\ref{dyn-chi})
can then
be recognized as the usual electromagnetic response function
(${\bf J}$-response to a spatially modulated ${\bf A}$-field)
in the zero-frequency limit.
This is one of the central results of this work.

%
Its relevance to the calculation of the macroscopic electromechanical tensors can
be appreciated by looking at the lowest terms in its small-$q$ expansion.
At zero-th order in $q$ we have
\begin{equation}
\Delta \chi_{\alpha \beta}({\bf q}=0) = 2 \Delta
   g_{\alpha \beta}({\bf q} =0) + \frac{N}{\Omega} \delta_{\alpha \beta},
\end{equation}
By invoking the $f$-sum rule, one can show that the result vanishes,
consistent with expectations: As we said, the zero-th order in $q$
corresponds to a rigid translation, which should not produce any macroscopic
electronic current \emph{in the reference frame that moves with the
crystal}.
Similarly, this can be regarded as a manifestation of the gauge invariance
of electromagnetism in the context of macroscopic electromechanical response
properties.

The first order in $q$ also vanishes, again as a consequence of
gauge invariance. Physically, one can show that the ${\bf q}$-derivative of
$\Delta \chi_{\alpha \beta}({\bf q})$ describes the
${\bf J}$-response to a static ${\bf B}$-field, or equivalently
the ${\bf M}$-response (${\bf M}$ is the orbital magnetization)
to a static ${\bf A}$-field; both are forbidden in insulators,
and only allowed in certain categories of metals in a transport regime.~\cite{zhong-16}
This unambiguously proves that the contribution of
the gauge fields, via the dynamical term $\Delta \chi_{\alpha \beta}$, to the
macroscopic
%
piezoelectric tensor identically vanishes,
and can be regarded as providing a formal proof (\emph{a posteriori}) that the
metric tensor approach of Hamann {\em et al.}~\cite{hamann-metric}
rests on firm theoretical grounds.

\subsubsection{Relationship to orbital magnetism}

The interesting physics, in our present context, occurs at second order in $q$.
By using Eq.~(\ref{chi-expq}) and substituting Eqs.~(\ref{chitot})
and (\ref{dyn-chi}), we can write the gauge-field
contribution to the bulk flexoelectric tensor as
\begin{equation}
\Delta \mu_{mn,kl} = -\frac{1}{2}
  \frac{ \partial^2   \Delta \chi_{mn}({\bf q}) }{\partial q_k \partial q_l}\Big|_{{\bf q}=0} =
  -\frac{ \partial^2 \Delta g_{mn}({\bf q}) }{\partial q_k \partial q_l}\Big|_{{\bf q}=0}.
 \mylabel{eq-j2}
\end{equation}
To see that this expansion term is directly related to orbital magnetism
(earlier derivations were reported by Vignale (PRL 1991)
and Mauri and Louie~\cite{Mauri/Louie:96}),
define the magnetic susceptibility tensor as
\begin{equation}
M_\alpha = \chi^{\rm mag}_{\alpha \beta} B_\beta,
\end{equation}
(${\bf M}$ and ${\bf B}$ are the magnetization and the magnetic field, respectively),
which for a monochromatic ${\bf A}$-field implies 
(${\bf J} = -\bm{\nabla} \times {\bf M}$, and ${\bf B} = \bm{\nabla} \times {\bf A}$)
that the magnetically induced current density is
\begin{equation}
J_m =  \epsilon^{ml \alpha} q_l \chi^{\rm mag}_{\alpha \beta}  \epsilon^{\beta kn} q_k A_n.
\mylabel{J-mag}
\end{equation}

Now observe that, in our context, the vector potential is the time derivative
of the displacement field, and that the polarization is the time derivative of 
the current. By taking the time integral on both sides of Eq.~(\ref{J-mag}), and by
recalling Eq.~(\ref{chi-uP}) we have, then 
\begin{equation}
\Delta \chi_{mn}({\bf q}) \sim \epsilon^{ml \alpha} q_l \chi^{\rm mag}_{\alpha \beta}  
   \epsilon^{\beta kn} q_k.
\end{equation}
Now, we can derive both sides twice with respect to ${\bf q}$, which leads to
\begin{equation}
 \Delta \mu_{mn,kl} = \frac{1}{2} \sum_{\alpha \beta}
  ( \epsilon^{\alpha m k} \epsilon^{\beta n l} + \epsilon^{\alpha m l} \epsilon^{\beta n k} ) \chi^{\rm mag}_{\alpha \beta}.
  \mylabel{j2-chi}
\end{equation}
In the special case of a solid with cubic symmetry, where $\chi^{\rm mag}_{\alpha \beta} = \chi^{\rm mag} \delta_{\alpha \beta}$,
the above expression can be simplified by using
$$\sum_\alpha \epsilon^{\alpha mk} \epsilon^{\alpha nl} = \delta_{mn} \delta_{kl} - \delta_{ml} \delta_{nk},$$
which leads to
\begin{equation}
  \Delta \mu_{mn,kl} = \frac{\chi^{\rm mag}}{2}
  (2 \delta_{mn} \delta_{kl} - \delta_{ml} \delta_{nk} - \delta_{mk} \delta_{nl} ) .
\end{equation}
Thus, in a cubic solid only two independent combinations of indices yield a nonzero value,
\begin{equation}
 \Delta \mu_{11,22} =  \chi^{\rm mag}, \qquad
 \Delta \mu_{12,12} = - \frac{\chi^{\rm mag}}{2}.
 \mylabel{delmu-cub}
\end{equation}

The fact that the flexoelectric response involves a contribution that
is exactly proportional to the diamagnetic susceptibility may appear
surprising at first sight, as this result combines two material properties
that are, at first sight, completely unrelated.
Yet, by recalling the equivalence between rotations and magnetism discussed
in Section~\ref{sec:larmor}, the above result, which is one of the key messages of this work,
becomes reasonable: Certain components
of the strain-gradient tensor involve gradients of the local rotation. A uniform
rotation, in turn, produces an orbital magnetization, ${\bf M}$; then, a rotation
\emph{gradient} that is applied adiabatically to the crystal produces a macroscopic
current (recall the relationship from electromagnetism ${\bf J}= -\nabla \times {\bf M}$)
that, integrated over time, yields a macroscopic polarization.

To summarize this long Section, we have achieved a decomposition of
the electronic flexoelectric tensor into two physically distinct
terms,
\begin{equation}
\bm{\mu} = \bar{\bm{\mu}} + \Delta \bm{\mu}.
\end{equation}
At this point, we are left with the obvious questions of whether the two contributions
$\bar{\bm{\mu}}$ and $\Delta \bm{\mu}$ are separately measurable and,
if yes, of how they should be treated in the
perspective of comparing the results to the experiments.
To provide a reliable answer, however, one needs to account for the
surface contributions along side the bulk ones, as we
know
that the two form an undissociable entity in the context of the flexoelectric response.
We shall discuss this topic in the following section.



\section{Microscopic polarization response and surface contributions}


To quantify the surface contributions to the flexoelectric response of a finite
object, we need to adapt the theory developed in the previous Section to
the calculation of the \emph{microscopic} polarization response to a deformation.
(The physical properties of the surface substantially differ from those of the
bulk, thus requiring a spatially resolved description.)
In particular, we shall be concerned with the response functions
$\bar{\chi}_{\alpha \beta}^{\bf q} ({\bf G})$ and $\Delta {\chi}_{\alpha \beta}^{\bf q} ({\bf G})$,
which we define by generalizing their macroscopic counterparts, Eq.~(\ref{stat-chi}) and Eq.~(\ref{dyn-chi}),
as follows,
\begin{eqnarray}
\bar{\chi}^{\bf q}_{\alpha \beta}({\bf G}) &=& 2 \int [d^3k] \sum_{v}
  \langle \phi_{v\bf k} |  \hat{\mathcal{J}}_{ \alpha \bf k,G + q} | \delta \phi_{v\bf k}^{(\lambda_\beta)}({\bf q})
\rangle, \nonumber \\ 
\Delta {\chi}^{\bf q}_{\alpha \beta}({\bf G}) &=& 2 \int [d^3k] \sum_{v}
  \langle \phi_{v\bf k} |   \hat{\mathcal{J}}_{ \alpha \bf k,G + q}  |  \phi_{v\bf k}^{(\dot \lambda_\beta)}({\bf q})
\rangle \nonumber \\
 && + \delta_{\alpha \beta} n_{\rm el}^{(0)} ({\bf G}).
\end{eqnarray}
The only difference with respect to the previous formulas is that the polarization response is now
calculated at ${\bf G+q}$, where ${\bf G}$ is a vector of the reciprocal-space Bravais lattice.
%
[Note that the average electron density, $n_{\rm el}({\bf G}=0)$, corresponds 
to $N/\Omega$, consistent with the macroscopic formula, Eq.~(\ref{dyn-chi}).]
Of course, the above expressions include the macroscopic
response defined earlier as a special case,
\begin{eqnarray}
\chi_{\alpha \beta}^{\bf q} ({\bf G}=0) &=& \chi_{\alpha \beta}({\bf q}),
\end{eqnarray}
where $\chi$ stands for either $\bar{\chi}$ or $\Delta \chi$.

\subsection{The role of the gauge fields}

To make a more direct connection with the existing treatments of the surface problem,
we shall assume a slab geometry henceforth, with the surface normal oriented along $x$,
and periodic boundary conditions in the $yz$ plane.
As in earlier works, we shall adopt open-circuit electrical boundary conditions
along $x$, as appropriate for a slab with free surfaces, and focus our
attention on the total \emph{open-circuit voltage} that is linearly induced
by a strain-gradient deformation.
To determine such ``flexovoltage''~\cite{artcalc} response we need
the induced electrostatic potential and this, in turn, is uniquely given
(modulo an irrelevant global constant) by the charge-density response of
the system to the perturbation.
This observation makes the analysis of a finite object conceptually simpler
than that of a bulk crystal -- the explicit inclusion of the boundaries
allows us to study the charge rather than the polarization, which is much
easier to define and calculate.

The charge response functions that are associated with the static and
dynamic terms can be written as minus the divergence of the polarization
response, which in reciprocal space can be written as ($\rho$ stands for
$\bar{\rho}$ or $\Delta \rho$, and $\chi$ for either $\bar{\chi}$ or $\Delta \chi$)
\begin{eqnarray}
\rho_\beta^{\bf q}({\bf G}) &=& -i \sum_\alpha (G_\alpha + q_\alpha) \chi^{\bf q}_{\alpha \beta}({\bf G}), 
\end{eqnarray}
Crucially, the
dynamic gauge-field contribution
to the charge-density response vanishes identically,
\begin{equation}
\Delta \rho_\beta^{\bf q}({\bf G}) = 0.
\mylabel{norho}
\end{equation}
This result may appear surprising at first sight, but it is really
a simple consequence of time-reversal symmetry: In absence of spin-orbit
coupling, a vector potential
field applied to the orbital degrees of freedom produces, in the linear regime,
a divergenceless circulating current, which does not alter the ground-state
electron density.
Still, the situation is paradoxical in light of the results of the previous Section:
How can we reconcile the irrelevance of the gauge fields for the electromechanical
response of a slab, clearly stated by Eq.~(\ref{norho}), with their nonvanishing
contribution to the bulk flexoelectric tensor, as expressed by Eq.~(\ref{j2-chi})?
The answer, as we anticipated at the end of the previous Section, resides
in the presence of surface contributions to the overall flexo-response of a
slab that are equal in magnitude and opposite in sign to $\Delta \bm{\mu}$,
leading to an exact cancellation of their combined effect.

\begin{figure}
\begin{center}
\includegraphics[width=2.8in]{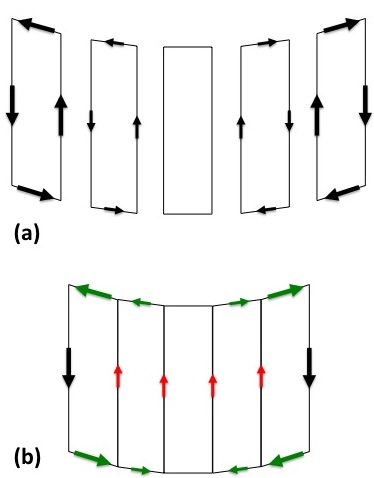}
\end{center}
\caption{ \label{sketch} Dynamical vector-potential 
contributions to the polarization of a slab
subjected to a shear strain gradient. (a) Decomposition into circulating
currents withing segments; (b) Decomposition between bulk and surface contributions.}
\end{figure}

To prove that such a cancellation indeed occurs, it suffices to review
Section~\ref{sec:larmor}, where the equivalence between a uniform rotation of
the sample and an effective orbital magnetic field is established; we shall see
that this result can quantitatively explain both the bulk and surface contributions
of the gauge fields to the flexoelectric response.
It is convenient, to that end, to introduce
a quantity ${\bf T}({\bf r})$ corresponding to
the time integral of the orbital magnetization,
\begin{equation}
{\bf T}({\bf r}) = \int_0^t {\bf M}({\bf r},t) dt,
\label{PCT}
\end{equation}
and since ${\bf J}=\bm{\nabla} \times{\bf M}=d{\bf P}/dt$, it follows that
\begin{equation}
{\bf P}({\bf r}) = - \bm{\nabla} \times {\bf T}({\bf r}).
\label{TCT}
\end{equation}
Loosely speaking, $\bf T$ can be thought of as a kind of
electric toroidization.
In the linear-response regime, the dynamic gauge-field
term in the Hamiltonian produces a ${\bf T}$-field whose amplitude
is proportional to the local rotation of the sample (we neglect the
spatial dispersion of the orbital diamagnetic response, which is
irrelevant in the context of the present discussion)
with respect to the unperturbed configuration [recall Eq.~(\ref{larmor})],
\begin{equation}
{\bf T}({\bf r}) = - 2\chi^{\rm mag} \bm{\theta} ({\bf r}).
\end{equation}
Now consider a displacement field of the type
\begin{equation}
u_y({\bf r}) = \frac{\eta}{2} x^2,
\end{equation}
corresponding to a uniform shear strain gradient applied to the slab,
as illustrated in Fig.~\ref{sketch}.
The rotation angle is given by
$\theta_z = \eta x /2$, and its curl is readily given by
$\bm{\nabla} \times \bm{\theta} = -\hat{\bf y} /2$;
Eqs.~(\ref{PCT}-\ref{TCT}) then yield
a contribution to the bulk flexoelectric response equal to $\Delta P_y = \eta \chi^{\rm mag}$,
consistent with Eq.~(\ref{delmu-cub}).

To gain a more intuitive insight into this result, one can regard the strain-gradient
deformation of the slab as a piecewise shear [Fig.~\ref{sketch}(a)], which we suppose to be
uniform within individual segments. (The realistic physical picture is then recovered
upon reducing the segment length to zero.)
The rotation of each segment is associated with a circulating surface polarization (black arrows),
and as the rotation amplitude linearly increases along $x$, the contribution of the facets
that lie next to each other (i.e. within the interior of the slab) does not cancel out;
on the contrary, they result in a uniform ${\bf P}$ [red arrows in Fig.~\ref{sketch}(b)].
In principle, a bulk polarization would result in a net surface charge; however,
the cartoon of Fig.~\ref{sketch}(a) clearly illustrates why here this is not the case.
Indeed, in addition to the aforementioned bulk effect, there is also a polarization
that develops at the outer surfaces of the segments [green arrows in Fig.~\ref{sketch}(b)].
Such a surface polarization is oriented in-plane, and linearly increases along the same direction $x$.
This polarization field yields (recall $\rho = - \bm{\nabla} \cdot {\bf P}$)
a uniform, net surface charge that exactly cancels the contribution of the bulk,
thereby settling the paradox that we described at the
beginning of this Section.

Summarizing the above, there are two equally valid ways to understand the
gauge-field contribution to the polarization field induced by a deformation,
which are illustrated by the two panels in Fig.~\ref{sketch}. We can think
of it either [panel (a)] as the sum of local circulating currents that arise because
individual segments of the slab undergo a local rotation with respect to the
original configuration; or [panel (b)], as in earlier treatments of the flexoelectric
problem, as a sum of bulk and surface contributions.
In either case, the overall sum yields a vanishing charge density, and is therefore
irrelevant in the context of an electrical measurement.
The conclusion is that we can discard the ``dynamical'' contributions to the
flexoelectric effect altogether, and build a predictive theory of the
electromechanical response based on the static contribution only.

\subsection{Connection to the existing theory of flexoelectricity}

%
%

Our next task is to clarify how all of the above relates to
the calculations of flexoelectricity that have recently been
reported.\cite{artcalc,Hong-11,Hong-13} These
previous works based their analysis on the microscopic response functions $P_{\alpha,\kappa \beta}^{\bf q}({\bf r})$,
which are defined as the $\alpha$ component of the polarization response, calculated in the laboratory frame, to a
monochromatic displacement of the sublattice $\kappa$ along the Cartesian direction $\beta$.
To summarize this approach,
it is useful to begin by considering the
sum of the above sublattice displacements, which corresponds to the polarization
response to an acoustic phonon in the laboratory frame. Following Ref.~\onlinecite{artgr}
we shall define
\begin{equation}
P_{\alpha \beta}^{\bf q}({\bf r}) = \sum_\kappa P_{\alpha,\kappa \beta}^{\bf q}({\bf r}).
\end{equation}
By taking into account the transformation properties of the current density between the laboratory
and the curvilinear frame in the linear regime of small deformations, one can then write the following
relationship,
\begin{equation}
P_{\alpha \beta}^{\bf q}({\bf r}) = \delta_{\alpha \beta} \rho^{(0)} ({\bf r}) + \chi_{\alpha \beta}^{\bf q} ({\bf r}),
\mylabel{p-chi}
\end{equation}
where the last term on the rhs is defined as the Fourier transform of the microscopic response function
$\chi_{\alpha \beta}^{\bf q} = \bar{\chi}_{\alpha \beta}^{\bf q} + \Delta \chi_{\alpha \beta}^{\bf q}$,
\begin{equation}
\chi_{\alpha \beta}^{\bf q} ({\bf r}) = \sum_{\bf G} \chi_{\alpha \beta}^{\bf q} ({\bf G}) e^{i {\bf G\cdot r}},
\end{equation}
and $\rho^{(0)} ({\bf r})$ is the ground-state charge density, inclusive of the nuclear point charges.

To access the macroscopic electromechanical properties of the system, a long-wave
decomposition is performed,~\cite{artgr}
\begin{equation}
P_{\alpha \beta}^{\bf q}({\bf r}) =  P_{\alpha \beta}^{(0)}({\bf r}) - iq_\gamma P_{\alpha \beta}^{(1,\gamma)}({\bf r}) -\frac{q_\gamma q_\delta}{2}
P_{\alpha \beta}^{(2,\gamma\delta)}({\bf r}) + \cdots,
\mylabel{p-exp}
\end{equation}
where the cell averages of the expansion terms yield the
electronic parts of the
macroscopic piezoelectric and flexoelectric tensors,
\begin{eqnarray}
-\frac{1}{\Omega} \int_{\rm cell} d^3 r P_{\alpha \beta}^{(1,\gamma)}({\bf r}) &=& e_{\alpha, \beta\gamma}, \\
\frac{1}{2 \Omega} \int_{\rm cell} d^3 r P_{\alpha \beta}^{(2,\gamma\delta)}({\bf r}) &=& \mu_{\alpha\beta,\gamma\delta}, \mylabel{mu-def}
\end{eqnarray}
consistent with the ${\bf q}$-expansion of the macroscopic $\chi_{\alpha \beta}({\bf q})$ tensors defined in
the previous Sections.
In fact, after observing that at ${\bf q}=0$ the microscopic polarization response function $\chi_{\alpha \beta}^{\bf q} ({\bf r})$
vanishes identically, one can use Eq.~(\ref{p-chi}) to directly relate the ${\bf q}$-expansion of the laboratory ${\bf P}$-response
to that of the curvilinear ${\bf P}$-response even at the microscopic
level \st{, e.g.} by writing
\begin{equation}
\chi_{\alpha \beta}^{\bf q} ({\bf r}) = i q_\gamma \chi_{\alpha \beta}^{(1,\gamma)} ({\bf r}) - q_\gamma q_\delta
\chi_{\alpha \beta}^{(2,\gamma\delta)}({\bf r}) + \cdots,
\end{equation}
and equating terms at each order in ${\bf q}$.

At this point, one would be tempted to proceed as in Ref.~\onlinecite{artgr}, and identify the first- and second-order
expansion terms as the microscopic polarization response to a uniform strain and to a strain gradient, respectively.
(This step was a crucial prerequisite to the calculation of the transverse components of the bulk flexoelectric tensor
that was performed in Ref.~\onlinecite{artgr}.)
This implies tentatively writing the induced polarization as
\begin{equation}
{\bf P}({\bf r}) = \varepsilon_{\beta\gamma} ({\bf r}) {\bf P}_{\beta \gamma}^{\rm U}({\bf r}) +
                   \frac{\partial \varepsilon_{\beta\gamma} ({\bf r})}{\partial r_\delta} {\bf P}_{\beta \gamma, \delta}^{\rm G}({\bf r}) + \cdots,
\mylabel{tentative}
\end{equation}
where $\varepsilon_{\beta\gamma} ({\bf r})$ is a spatially nonuniform
symmetric strain field, and ${\bf P}^{\rm U}$ and ${\bf P}^{\rm G}$
describe
the linear polarization response to a uniform (U) strain and to its gradient (G), respectively.
In Refs.~\onlinecite{artgr,artcalc} it was assumed that such response functions simply correspond to the
${\bf q}$-expansion terms of $P_{\alpha \beta}^{\bf q}({\bf r})$.
In light of the results of this work, however, an expression such as Eq.~(\ref{tentative}) is
physically problematic, as it implicitly assumes that the polarization response to a rigid translation or a
rotation of the crystal vanishes. While we know this to be true for translations, rigid rotations do contribute to
${\bf P}({\bf r})$ via the dynamic gauge-field terms discussed in the previous section.
As a consequence, we cannot identify $P_{\alpha,\beta \gamma}^{\rm U}({\bf r})$ with
either $\chi_{\alpha \beta}^{(1,\gamma)} ({\bf r})$ or, equivalently, with $-P_{\alpha \beta}^{(1,\gamma)}({\bf r})$:
$P_{\alpha,\beta \gamma}^{\rm U}$ is symmetric with respect to $\beta\gamma$ by construction, while
the other two functions implicitly contain an antisymmetric contribution that is mediated by the
gauge-field rotation response.

As we anticipated in the previous Section, an elegant solution to this problem consists
in dropping the dynamic gauge-field response altogether, and writing the theory in
terms of the static response function $\bar \chi$ only.
The latter enjoys a ${\bf q}$-expansion analogous to that of the total $\chi$,
\begin{equation}
\bar{\chi}_{\alpha \beta}^{\bf q} ({\bf r}) = i q_\gamma \bar{\chi}_{\alpha \beta}^{(1,\gamma)} ({\bf r}) - q_\gamma q_\delta
\bar{\chi}_{\alpha \beta}^{(2,\gamma\delta)}({\bf r}) + \cdots,
\end{equation}
with the key advantage that the first-order term is now symmetric under $\beta\gamma$ exchange.
This formally justifies
the use of Eq.~(\ref{tentative}), together with the definitions
\begin{eqnarray}
P_{\alpha,\beta \gamma}^{\rm U}({\bf r}) &=& \bar{\chi}_{\alpha \beta}^{(1,\gamma)} ({\bf r}), \mylabel{pu-chibar} \\
P_{\alpha \delta,\beta \gamma}^{\rm G}({\bf r}) &=& \bar{\chi}_{\alpha \beta}^{(2,\gamma\delta)}({\bf r}) +
 \bar{\chi}_{\alpha \gamma}^{(2,\beta\delta)}({\bf r}) - \bar{\chi}_{\alpha \delta}^{(2,\beta\gamma)}({\bf r}), \mylabel{pg-chibar}
\end{eqnarray}
where we have operated the standard permutation of indices on the rhs of Eq.~(\ref{pg-chibar})
in order to move from a ``type-I'' (second gradient of the displacement field) 
to a ``type-II'' (first gradient of the symmetrized strain tensor) representation
of the strain-gradient tensor.~\cite{artlin,chapter-15}
This way, we can connect the present analytical results with the existing theory of
the flexoelectric response. Most importantly, this allows us to formally reconcile
the existing calculations of the bulk flexoelectric tensor, which were based on an
analysis of the charge-density response in a supercell geometry,~\cite{artcalc} with
the more fundamental current-response theory that we have developed in this work.

\subsection{Calculation of the transverse components}


\begin{figure}
\begin{center}
\includegraphics[width=2.8in]{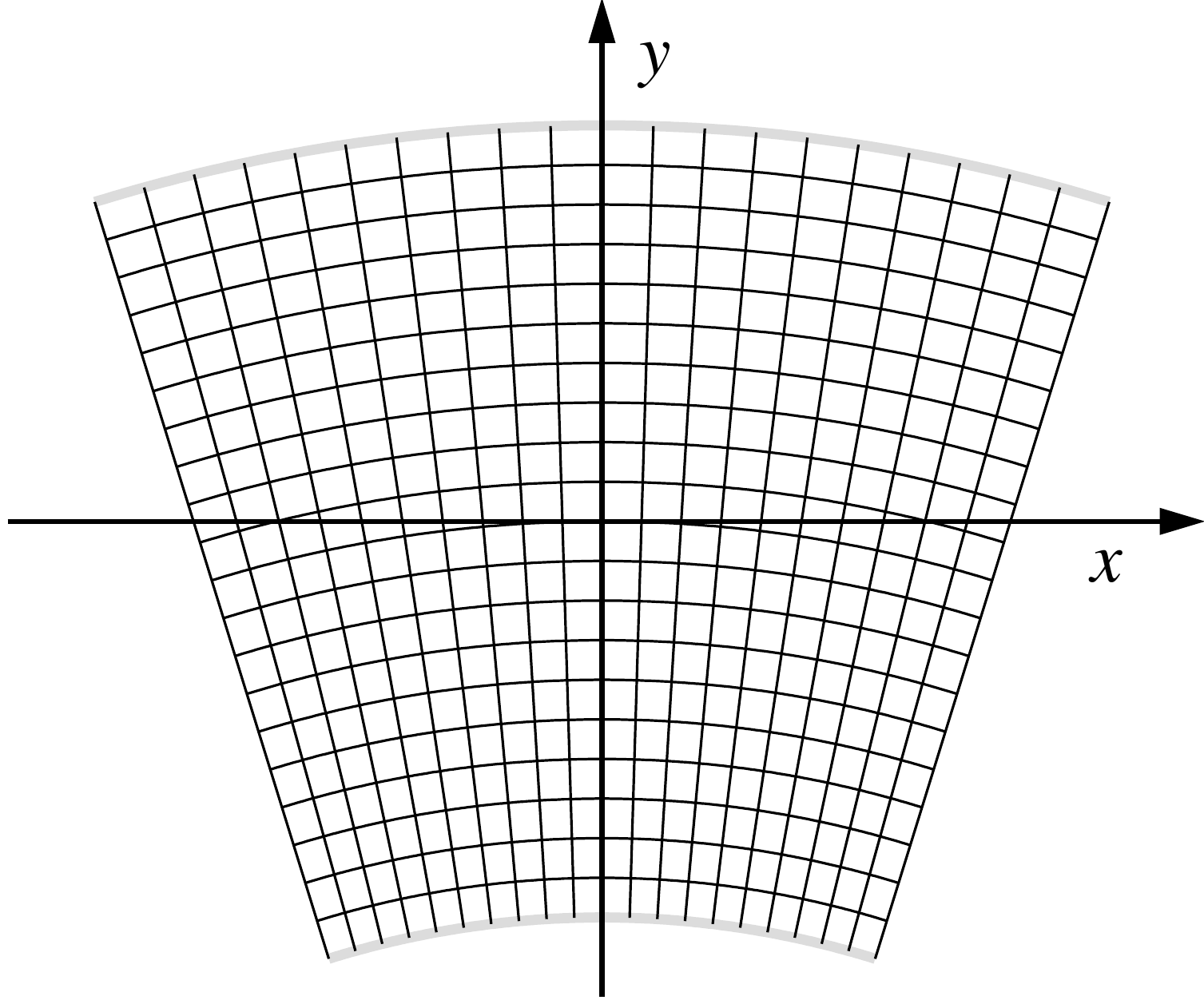}
\end{center}
\caption{ \label{bend} Schematic illustration of the flexural deformation of a slab. 
Thick arrows indicate the Cartesian axes, thick gray curves indicate the slab surfaces.}
\end{figure}

To illustrate the above arguments, it is useful to explicitly work out the
example of a symmetric slab, finite in the $y$ direction,
subjected to a transverse (i.e., flexural) strain gradient
deformation via the displacement field
\begin{eqnarray}
u_x &=& \eta xy, \\
u_y &=& -\frac{\eta}{2} x^2
\end{eqnarray}
(see Fig.~\ref{bend}).
The symmetric strain tensor has only
one non-zero entry,
\begin{equation}
\varepsilon_{xx} = \frac{\partial u_x}{\partial x} = \eta y,
\end{equation}
indicating a linear increase of the transverse component along the normal
to the slab surface, i.e., a constant strain-gradient field of the type
\begin{equation}
\varepsilon_{xx,y} = \frac{\partial \varepsilon_{xx}}{\partial y} = \eta.
\end{equation}
By using Eq.~(\ref{tentative}), we can readily write the resulting polarization
field (within the linear approximation and discarding higher-order gradient effects)
as
\begin{equation}
\frac{\partial {\bf P}({\bf r})}{\partial \eta} = y  {\bf P}_{xx}^{\rm U}({\bf r}) +
                   {\bf P}_{xx, y}^{\rm G}({\bf r}).
      \mylabel{p-bend}
\end{equation}

To move further it is convenient to operate, as customary, a macroscopic averaging
procedure on the ${\bf P}^{\rm U}$ and ${\bf P}^{\rm G}$ functions in order to filter out the irrelevant
oscillations on the scale of the interatomic spacings.
This way, the in-plane spatial resolution is completely
suppressed, leaving response functions that depend
on $y$ only.
%
Note that, by symmetry, the induced polarization can only have nonzero
$y$ components,
\begin{equation}
\frac{\partial P_y(y)}{\partial \eta} = y  P_{y,xx}^{\rm U}(y) +
                   P_{yy,xx}^{\rm G}(y),
                   \mylabel{py-bend}
\end{equation}
and that since the bulk is nonpiezoelectric the first function on the rhs, $P_{y,xx}^{\rm U}(y)$,
can only be nonzero near the surface.
These observations allow one to conclude
that the induced \emph{surface charge}, $\sigma$ is uniquely determined by the second term
on the rhs of Eq.~(\ref{py-bend}); in the limit of a thick slab, we can then write
\begin{equation}
\frac{\partial \sigma}{\partial \eta} = \frac{\bar{\mu}_{\rm T}}{\epsilon_\infty},
\mylabel{sigma}
\end{equation}
where
\begin{equation}
\bar{\mu}_{\rm T} = 2 \bar{\mu}_{12,12} - \bar{\mu}_{11,22}
\end{equation}
is the transverse component of the ``barred'' (no gauge fields) bulk flexoelectric tensor
in type-II form, and $\epsilon_\infty$ is the bulk relative permittivity at the clamped-ion
level.

We stress that the above arguments, linking $\bar{\mu}_{\rm T}$ to the
surface charge $\sigma$, and hence to the macroscopic electric field
that develops in the interior of the slab upon bending, $E_y = -\sigma/ \epsilon_0$
($\epsilon_0$ is the vacuum permittivity), hold under the hypothesis
that the effect of rotations is excluded from Eq.~(\ref{tentative}), which can
only be justified if the gauge-field contribution is excluded from \emph{both}
bulk and surface contributions to the overall flexoelectric response.
This means that the explicit calculation of the bulk flexoelectric tensor
of SrTiO$_3$ that was carried out in in Ref.~\onlinecite{artcalc} really
concerned $\bar{\bm{\mu}}$, and not the total tensor $\bm{\mu}= \bar{\bm{\mu}} + \Delta \bm{\mu}$.
We believe that the former quantity, $\bar{\bm{\mu}}$, given its more direct
relationship to the charge-density response of the system, is physically
more meaningful than $\bm{\mu}$, and should be preferred to the latter
when reporting the results of first-principles calculations.

\subsection{A simple example}

Consider a simple cubic lattice made of spherical, closed-shell atoms,
with a cell parameter that is sufficienty large as to avoid any direct
interaction between neighboring sites.
Such a crystal is, of course, unrealistic as there is no force
whatsoever keeping the atoms in place. Nevertheless, it is a useful toy
model to discuss some fundamental aspects of the flexoelectric response,
without the complications that characterize a real material.
This model was introduced in Ref.~\onlinecite{artgr} to illustrate some
subtleties related to surface contributions; here we shall
%
use it to illustrate the two alternative definitions of the
bulk flexoelectric tensor, either excluding ($\bar{\bm{\mu}}$) or including ($\bm{\mu}$)
the dynamic gauge-field response.

The basic quantities that define the model are: (i) the spherical charge
distribution $\rho_{\rm at}(r)$ of each isolated atom, and
(ii) the lattice parameter $a_0$. Then the charge density can be
readily written as
\begin{equation}
\rho({\bf r}) = \sum_{\bf R} \rho_{\rm at}(| {\bf r-R} |),
\end{equation}
where the sum runs over the Bravais lattice defined by $a_0$. To
calculate the flexoelectric tensor via the current-density response we
need the microscopic polarization field that is induced by the
displacement of an isolated atom.
As the atoms are spherical, there are no long-range electrostatic forces
involved, and since they are noninteracting, one can readily use
the transformation laws of the probability current to write
\begin{equation}
\mathcal{P}_{\alpha,\beta } ({\bf r}) =
\rho_{\rm at}(r) \delta_{\alpha \beta}
\end{equation}
where we have dropped the sublattice index $\kappa$ since we are dealing with
a single atom per unit cell.
This equation reflects the fact that the probability current associated
with an isolated spherical atom located at the origin and moving with
uniform velocity ${\bf v}$ is simply given by ${\bf v}$ times
the atomic charge density,
\begin{equation}
{\bf J}({\bf r}) = {\bf v} \rho_{\rm at}(r).
\end{equation}
Now, recall the definition of the flexoelectric tensor
given by Eq.~(\ref{mu-def}).
%
%
We have, for the three independent components,
\begin{equation}
\mu_{\rm L} = \mu_{\rm S} = \frac{Q}{2\Omega}, \qquad \mu_{\rm T} = -\frac{Q}{2\Omega},
\end{equation}
where $Q$ is the quadrupolar moment of the static atomic charge,
\begin{equation}
Q = \int d^3 r \rho_{\rm at}(r) x^2,
\end{equation}
and the longitudinal (L), transverse (T) and shear (S) components are
given by
$$\mu_{\rm L} = \mu_{11,11}, \qquad \mu_{\rm S} = \mu_{11,22},$$
\begin{equation}
\mu_{\rm T} = 2\mu_{12,12} - \mu_{11,22}.
\end{equation}
%

To calculate the ``revised'' version of the flexoelectric tensor we
need to calculate the gauge-field contribution, which is in turn given
by the macroscopic diamagnetic susceptibility via Eq.~(\ref{delmu-cub}),
$$\Delta \mu_{\rm L} = 0, \qquad \Delta \mu_{\rm S} = \chi^{\rm mag},$$
\begin{equation}
\Delta \mu_{\rm T} =-2 \chi^{\rm mag}.
\end{equation}
Given the noninteracting nature of the spherical atoms, we can
apply Langevin theory to calculate $\chi^{\rm mag}$,
\begin{equation}
\chi^{\rm mag} = \frac{Q}{2\Omega},
\end{equation}
which immediately yields
\begin{equation}
\mu'_{\rm L} = \mu'_{\rm T} = \frac{Q}{2\Omega}, \qquad \mu'_{\rm S} = 0.
\mylabel{muprim}
\end{equation}
Eq.~(\ref{muprim}) matches the conclusions of earlier works,
where the flexoelectric tensor components were inferred from the
behavior of the macroscopic electrostatic potential under a deformation.
(The interested reader can find a
detailed derivation in the Supplementary Note 1 of
Ref.~\onlinecite{artgr}, or in Ref.~\onlinecite{chapter-15}.)

%
%
%

By comparing the two ``versions'' of the flexoelectric tensor,
it is clear that the quantitative differences can be substantial, even in the trivially
simple case of the toy model described in this section.
Further work is needed to assess the impact of these effects on
the calculation of flexoelectricity in realistic materials.
In any case, the discussion presented should serve as a warning against
potential misunderstandings when interpreting the results of calculations
of flexoelectric responses.
%

\section{Discussion}

  It is important to stress that Eq.~(\ref{tentative}), together
  with the definitions of Eqs.~(\ref{pu-chibar}) and~(\ref{pg-chibar}), does not describe the total
  polarization response, but only a part of it. This part is enough for an
  exact description of electromechanical effects, as we have seen in the case
  of flexoelectricity.
One can wonder, however, whether there is any physical significance that can
be associated with the part that we have discarded from our analysis, i.e.,
the gauge-field contribution.
In this Section we shall briefly discuss this topic.

The connection of rotations and orbital magnetization has been noted
earlier in other contexts; for example, it plays an important role in the
theory of molecular $g$-factors.
Ceresoli and Tosatti~\cite{Ceresoli-02} (CT) have shown how such quantities can
be understood (and calculated from first principles) as the Berry phases
that the wavefunctions accumulate in the course of a rotation of
the molecule around its center of mass.
It is interesting to analyse their approach in some detail, in order to
show its strong relationship to the topics of the present work.

CT base their formalism on the electronic ground state of an isolated molecule,
whose rotation state about the $z$ axis
is measured by an angle, $\theta$.
The instantaneous ground state of the molecule is defined by the lowest $N$
eigenstates of the Hamiltonian, which depend parametrically on $\theta$,
\begin{equation}
\hat{\mathcal{H}}(\theta) |\psi_n(\theta)\rangle = \epsilon_n |\psi_n(\theta)\rangle.
\end{equation}
($\epsilon_n$ does not depend on $\theta$, as the energy of the system is
invariant upon rotations.) Then, by discretizing the $[0,2\pi]$ interval
into $M$ equally spaced points $\theta_i$, one can write the Berry phase
corresponding to a complete cycle as
\begin{equation}
\gamma \simeq -{\rm Im} \, \log \prod_{i=1,M} \det \, {\bf S}(\theta_i,\theta_{i+1}),
\mylabel{Berry}
\end{equation}
where $S$ are $N\times N$ matrices,
\begin{equation}
S_{mn}(\theta_i,\theta_j) = \langle \psi_m(\theta_i) |\psi_n(\theta_j)\rangle,
\end{equation}
and we have enforced periodic boundary conditions on the wavefunction gauge,
\begin{equation}
|\psi_n(\theta_{M+1})\rangle = |\psi_n(\theta_1)\rangle.
\mylabel{pbc}
\end{equation}

Note that $\gamma$ is a well-defined physical observable in spite of the 
arbitrariness of the wavefunction phases,~\cite{resta-jpcm2000} and vanishes 
identically in the absence of an applied magnetic field.
The strategy taken by CT was to assume that a small uniform ${\bf B}$-field,
oriented along the rotation axis, was
applied in the calculation of the instantaneous ground states that define $\gamma$.
In particular, one can introduce the Berry curvature that is
associated with the two-dimensional parameter space $(B,\theta)$,
\begin{equation}
\Omega_{B\theta} = -2\, {\rm Im} \sum_n \langle \psi_n^{(B)} | \psi_n^{(\theta)} \rangle,
\end{equation}
where the superscripts indicate the first-order wavefunctions
with respect to either $B$ or $\theta$. These, in turn, can be written 
as sums over conduction states,
\begin{equation}
|  \psi_{n}^{(\lambda)} \rangle =
     \sum_{c}  | \psi_{c} \rangle \frac{ \langle \psi_{c} | \partial \hat{\mathcal{H}} / \partial \lambda  | \psi_{n} \rangle}
  {\epsilon_{n} - \epsilon_{c}},
\end{equation}
where $\partial \hat{\mathcal{H}} / \partial \lambda$ is, as usual, the variation of the Hamiltonian at 
linear order in the perturbation parameter.
It is easy then to show~\cite{resta-jpcm2000} that, at linear 
order in $B$, $\gamma$ is the flux of $\Omega_{B\theta}$ through the 
rectangle spanned by $B$ and $2\pi$, 
\begin{equation}
\gamma = 2\pi B \Omega_{B\theta}.
\end{equation}

In order to recast the above result into the formalism developed in this work,
we shall choose an electromagnetic gauge 
for the vector potential such that
\begin{equation}
{\bf A} = \frac{1}{2} {\bf B} \times {\bf r},
\end{equation}
where ${\bf B}= (0,0,B)$ and the coordinate origin coincides with the rotation axis of the molecule.
($\gamma$, of course, does not depend on the electromagnetic gauge; the above choice has been made in order to facilitate the analytic derivations that follow.)
Then, the first-order Hamiltonian with respect to the external ${\bf B}$ 
is
\begin{equation}
\hat{\mathcal{H}}^{(B)} = \frac{1}{2} \hat{z} \cdot \int {\bf r} \times \hat{\bm{\mathcal{J}}}({\bf r}) d^3 r,
\end{equation}
where $\mathcal{J}({\bf r})$ is the current-density operator in the Cartesian frame,
and $\hat{z}$ is a unit vector oriented along $z$.

One can then write
\begin{equation}
\langle \psi^{(B)}_n |  \psi^{(\theta)}_n \rangle = -\frac{i}{2}  \hat{z} \cdot
 \int d^3 r \,  {\bf r} \times  \langle \psi_{n} | \hat{\bm{\mathcal{J}}}({\bf r}) |  \delta \psi^{(\theta)}_n \rangle,
\end{equation}
where $|  \delta \psi^{(\theta)}_n \rangle$ is the adiabatic counterpart of the first-order wavefunction
$|  \psi^{(\theta)}_n \rangle$.
We can recognize, in the integral, the microscopic current-density field that is induced
by a uniform rotation of the molecule,
\begin{equation}
  \frac{\partial {\bf J}({\bf r})}{\partial \dot{\theta}} = 2 \, {\rm Re} \sum_n \langle \psi_{n} | \hat{\bm{\mathcal{J}}}({\bf r})
       |  \delta \psi^{(\theta)}_n \rangle.
\end{equation}
Then $\gamma$ can be readily rewritten, in the linear regime, as
\begin{equation}
\gamma = \pi B \hat{z} \cdot \int {\bf r} \times \frac{\partial {\bf J}({\bf r})}{\partial \dot{\theta}} d^3 r,
\end{equation}
i.e. it is proportional to the $z$-component of the electronic magnetic moment, ${\bf m}$, that is associated 
with the rotation,
\begin{equation}
\gamma = \pi  B \, \frac{\partial m_z}{\partial \dot{\theta}}.
\end{equation}
This also implies that 
\begin{equation}
\Omega_{B\theta}=\frac{1}{2}\,\frac{\partial m_z}{\partial \dot{\theta}}.
\end{equation}

To summarize, $\gamma$ tells us the electronic contribution to
the magnetic moment associated with the rotation of the molecule, which
could be combined with the trivial contribution from the nuclear motion
to compute the $g$-factor of the molecule as a whole.
Interestingly, though, the same $\gamma$ is also closely related
the the magnetic susceptibilty of the static molecule. In particular,
an earlier work~\cite{cybulski-94} demonstrated that
the quantity we call $\gamma$ corresponds to
the \emph{paramagnetic} part of the susceptibility of the molecule.
The theory developed here nicely fits with this result.

To see this, note that in the theory of molecular magnetic
susceptibility, the ``diamagnetic'' contribution is defined such
that it is given by the second moment of the ground-state electronic
density, and the ``paramagnetic part'' is defined as the remainder.
As we have discussed in Sec.~\ref{sec:larmor}, a uniform rotation at a 
frequency $\omega$ produces, in the rotating frame that is rigid
with the molecule, the same effects (at linear order) as a uniform
$B$-field, i.e., the sum of the diamagnetic and paramagnetic
pieces just discussed.
To get the total moment in the laboratory frame, as reflected in
$\gamma$, we have to add to this a trivial piece coming from
the rigid rotation of the ground-state electronic cloud, which
is just minus the diamagnetic contribution to the
susceptibility. Thus, it follows that $\gamma$
corresponds precisely to the paramagnetic part of the magnetic
susceptibility of the molecule.

Of course, the case of a molecule is relatively simple to deal with. Being  
an isolated object, it does not present serious technical issues 
no matter how the calculation is carried out (either by using the 
Ceresoli and Tosatti approach, or the linear response to $B$ as
discussed in the above paragraphs).
It would be interesting, however, to explore these ideas in the
case of extended solids, where orbital magnetic effects associated
with zone-center optical phonons have received some attention in the
past. In an infinite crystal, a finite magnetic field
(which CT used for calculating $\gamma$ via the Berry phase approach) 
is far less obvious to apply, and our linear-response strategy may
prove handy.
We shall leave this interesting topic for future investigations.

\section{Conclusions and outlook}

In summary, we have established a full-fledged quantum theory of
inhomogeneous mechanical deformations, by working within a 
linear-response density-functional framework.
An intimate and unsuspected connection to orbital magnetism has 
emerged, where the latter naturally enters as a consequence of a 
dynamically applied deformation of the crystal. 
This effect produces a contribution to the bulk flexoelectric coefficient 
that corresponds to the orbital magnetic susceptibility of the material.

An obvious question that may be asked is whether this unusual interplay
of elasticity and magnetism can lead to interesting new physics, beyond
the topics that we discussed in this work, in terms of
experimentally measurable effects.
We believe that the best candidates may be magnetic materials
in a proximity of a phase transition to a ferromagnetic state,
where the susceptibility peaks to huge values. 
However, ferromagnetism only occurs in presence of spins, and whether
deformations affect the spin degree of freedom in the same way we have
shown for the orbital ones, still remains to be seen. 
Interest in this mechanism has been growing in the past few years,
with the proposal that surface acoustic waves may be used to manipulate
the magnetic state of nanoparticles.~\cite{chudnovsky}
Thus, we regard this as a stimulating avenue for future research.

\section*{Acknowledgments}

We acknowledge the support of Ministerio de Econom\'ia, Industria y Competitividad (MINECO-Spain) through Grants MAT2016-77100-C2-2-P and SEV-2015-0496, Generalitat de Catalunya through Grant 2017 SGR1506, and Office of Naval Research (ONR) through Grant N00014-16-1-2951. This project has received funding from the European Research Council (ERC) under the European Union’s Horizon 2020 research and innovation programme (Grant Agreement No. 724529). 

\appendix

\section{Derivation of the curvilinear-frame Schr\"odinger equation}

\mylabel{app:a}

In this Appendix we shall back up the results of Sec.~\ref{sec:tdse}
with a more detailed derivation.

\subsection{Potential term}

The potential $V({\bf r},t)$ generally contains contributions from the
external potential of the nuclei, plus self-consistent Hartree and
exchange and correlation terms.
As in this manuscript we have assumed
an all-electron framework, the external
potential is that of the nuclear point charges.
Thus, $V({\bf r},t)$ reduces to an electrostatic (Hartree) term plus the
exchange and correlation potential,
\begin{equation}
V({\bf r},t) = V_{\rm H} ({\bf r},t) + V_{\rm XC} ({\bf r},t),
\end{equation}
where $V_{\rm H}$
is the solution of the following Poisson's equation,
\begin{equation}
\nabla^2 V_{\rm H}({\bf r}) = -4 \pi [n_{\rm el} ({\bf r}) - \rho_{\rm ion} ({\bf r})].
\end{equation}
Here $\rho_{\rm ion}$ is a sum of delta functions representing
the nuclei, $n_{\rm el} ({\bf r}) = |\psi({\bf r})|^2$
the electronic \emph{particle} density,
and $V_{\rm XC}$ is the functional derivative of the exchange and
correlation energy with respect to the electron density,
\begin{equation}
V_{\rm XC} ({\bf r}) = \frac{\delta E_{\rm XC}} {\delta n_{\rm el} ({\bf r})}.
\end{equation}

The transformation to a curvilinear coordinate system is relatively easy
for both the electrostatic and exchange and correlation terms.
First, we introduce the electron density
in the curvilinear frame,
\begin{equation}
\cur{n}_{\rm el}(\bm{\xi},t) = |\cur{\psi}(\bm{\xi},t)|^2 = h^{-1}(\bm{\xi},t) \, n_{\rm el} ({\bf r}(\bm{\xi},t),t),
\end{equation}
where we have used the shortcut $h= \det ({\bf h})$.
Then, the Poisson's equation in the curvilinear frame becomes
\begin{equation}
\partial_\alpha (h g^{\alpha \beta}  \partial_\beta V_{\rm H}) = -4 \pi (\cur{n}_{\rm el} - \cur{\rho}_{\rm ion}),
\end{equation}
where $\partial_\alpha = \partial/ \partial \xi_\alpha$ is the gradient
operator in $\bm{\xi}$-space, $\cur{\rho}_{\rm ion} = h^{-1} {\rho}_{\rm ion}$, and
\begin{equation}
g^{\alpha \beta}= ({\bf g}^{-1})_{\alpha \beta}
\mylabel{g-inv}
\end{equation}
is the inverse of the metric tensor.
This means that, from the point of view of the electrostatics, the
curvilinear frame is essentially equivalent to a Cartesian frame,
with one exception: the vacuum permittivity, $\epsilon_0$,
must be replaced with a (generally anisotropic) dielectric tensor,
$\bm{\epsilon}$, that in turn depends on the metric of the
deformation as
$\bm{\epsilon} = \epsilon_0 \sqrt{g} \, {\bf g}^{-1}$
where $g=\det{\bf g} = h^2$.
The exchange and correlation energy, at the level of the local density
approximation, can be written as
\begin{eqnarray}
E_{\rm XC} &=& \int d^3r  \, {n}_{\rm el}({\bf r}) \epsilon_{\rm XC}({n}_{\rm el}({\bf r})) \nonumber \\
           &=& \int d^3 \xi \, \cur{n}_{\rm el} (\bm{\xi})  \epsilon_{\rm XC}(h^{-1}(\bm{\xi}) \, \cur{n}_{\rm el}(\bm{\xi})),
           \mylabel{exc}
\end{eqnarray}
which leads to a straightforward expression for the potential.

\subsection{Kinetic term}

To derive the kinetic contribution to $\cop{H}$, one can start from the
Laplace-Beltrami operator and apply it to the curvilinear representation of
the wavefunction,
\begin{equation}
\nabla^2 \psi({\bf r},t) = 
\frac{1}{h} \partial_\alpha \left[ h g^{\alpha\beta} \partial_\beta \left( \frac{1}{\sqrt{h}} \cur{\psi}(\bm{\xi},t) \right) \right].
\end{equation}
After some tedious (but otherwise straightforward) algebra, one obtains
\begin{equation}
-\frac{1}{2} \nabla^2 \psi({\bf r},t) = \frac{1}{2\sqrt{h}} (\cop{p}_\beta -i \mathcal{A}_\beta ) g^{\beta \gamma} (\cop{p}_\gamma + i \mathcal{A}_\gamma )
\cur{\psi}(\bm{\xi},t),
\mylabel{kepart}
\end{equation}
where $\cop{p}_\alpha = -i \partial_\alpha$ is the canonical momentum operator in
$\bm{\xi}$-space, and $\mathcal{A}_\alpha$ is the auxiliary vector field defined in
Eq.~(\ref{mcA-def}).
This result almost exactly matches the expression derived by
Gygi~\cite{gygi-93}, except for a sign discrepancy in the contribution
of the ``vector potential'' $\mathcal{A}_\beta$ [see Eq. (7) therein].
One can then rewrite the kinetic contribution
%
to $\cop{H}$ as
\begin{eqnarray}
 \frac{1}{2} (\cop{p}_\beta -i \mathcal{A}_\beta ) g^{\beta \gamma} (\cop{p}_\gamma + i \mathcal{A}_\gamma )  &=&
 \frac{1}{2} \cop{p}_\beta g^{\beta \gamma} \cop{p}_\gamma + V_{\rm geom}(\bm{\xi}), \nonumber
\end{eqnarray}
where $V_{\rm geom}(\bm{\xi})$ corresponds to Eq.~(\ref{vgeom}). Thus, 
the auxiliary field $\bm{\mathcal{A}}$ does not really act
as a vector, but rather as a scalar potential.
%
Note that the $\bm{\mathcal{A}}$-field essentially
coincides (apart from a factor of 1/2) with the contracted Christoffel
symbol $\Gamma^\mu_{\mu \nu}$; thus, the operator $\hat{p}_\gamma + i
\mathcal{A}_\gamma$ can be thought as a sort of covariant derivative~\cite{gygi-93}
acting on the electronic wavefunctions.

\subsection{Time derivative}

Our starting point is
\begin{equation}
i \frac{\partial}{\partial t} \psi ({\bf r},t) = i \frac{\partial}{\partial t} 
\left[ \sqrt{| {\bf h}^{-1} ({\bf r},t) |} \cur{\psi} \left(\bm{\xi}({\bf r},t),t \right) \right],
\end{equation}
where $\bm{\xi}({\bf r},t)$ is the \emph{inverse} coordinate transformation
from ${\bf r}$-space to $\bm{\xi}$-space, and
\begin{equation}
h^{-1}_{\beta \gamma} = \frac{ \partial \xi_\beta ({\bf r},t) } {\partial r_\gamma}.
\end{equation}
Now observe that
\begin{equation}
\bm{\xi}({\bf r}(\bm{\xi},t),t) = \bm{\xi},
\end{equation}
which implies that 
\begin{equation}
\frac{ \partial \xi_\beta }{\partial t} \Big|_{\bf r} =
 -\frac{ \partial \xi_\beta }{\partial r_\gamma} \frac{ \partial r_\gamma }{\partial t} \Big|_{\bm{\xi}}.
\label{dxidt}
\end{equation}
(Note that the time derivative on the left-hand side has to be taken at fixed ${\bf r}$,
while the time derivative on the right-hand side is at fixed $\xi$ --
this is usually obvious, we made it explicit here to avoid possible sources of confusion.)

We shall derive things piece by piece. First, the derivative of the wavefunction,
\begin{equation}
\frac{\partial \cur{\psi}}{\partial t} \Big|_{\bf r} =
\frac{\partial \cur{\psi}}{\partial t} \Big|_{\bm{\xi}} + 
\frac{\partial \cur{\psi}}{\partial \xi_{\beta}}  \frac{\partial \xi_{\beta}}{\partial t} \Big|_{\bf r},
\end{equation}
by using Eq.~(\ref{dxidt}) becomes
\begin{equation}
\frac{\partial \cur{\psi}}{\partial t} \Big|_{\bf r} =
\frac{\partial \cur{\psi}}{\partial t} \Big|_{\bm{\xi}} - 
\frac{\partial \cur{\psi}}{\partial \xi_{\beta}} \frac{ \partial \xi_\beta }{\partial r_\gamma} \frac{ \partial r_\gamma }{\partial t} \Big|_{\bm{\xi}}.
\end{equation}
We can now insert an identity operator,
\begin{equation}
\frac{\partial \cur{\psi}}{\partial t} \Big|_{\bf r} =
\frac{\partial \cur{\psi}}{\partial t} \Big|_{\bm{\xi}} - 
  \frac{\partial \cur{\psi}}{\partial \xi_{\beta}} 
  \frac{ \partial \xi_\beta }{\partial r_\gamma} 
  \frac{ \partial \xi_\delta }{\partial r_\gamma}
  \frac{ \partial r_\lambda }{\partial \xi_\delta}
  \frac{ \partial r_\lambda }{\partial t} \Big|_{\bm{\xi}},
\end{equation}
and finally rewrite the above as
\begin{equation}
\frac{\partial \cur{\psi}}{\partial t} \Big|_{\bf r} =
\frac{\partial \cur{\psi}}{\partial t} \Big|_{\bm{\xi}} - A_\beta g^{-1}_{\beta \gamma}
  \frac{\partial \cur{\psi}}{\partial \xi_{\gamma}}.
\end{equation}
where $A_\beta$ is the effective vector potential of Eq.~(\ref{Adef}).
%
Second, the time derivative of the
volume prefactor reads as
\begin{equation}
\frac{\partial } {\partial t} \frac{1}{\sqrt{h}} \Big|_{\bf r} = 
  -\frac{1}{2 h\sqrt{h}} \frac{\partial h } {\partial t} \Big|_{\bf r} =  -\frac{1}{2 h\sqrt{h}} 
  \left( \frac{\partial h } {\partial t} \Big|_{\bm{\xi}} + 
  \frac{\partial h } {\partial \xi_\beta} \frac{\partial \xi_\beta} {\partial t} \Big|_{\bf r} \right).
\end{equation} 
By using again Eq.~(\ref{dxidt}), this leads to
\begin{equation}
\frac{\partial } {\partial t} \frac{1}{\sqrt{h}} \Big|_{\bf r} =  -\frac{1}{2 h\sqrt{h}} 
  \left( \frac{\partial h } {\partial t} \Big|_{\bm{\xi}} -
  \frac{\partial h }{\partial \xi_\beta} 
  \frac{\partial \xi_\beta}{\partial r_\lambda} \frac{\partial r_\lambda}{\partial t} \right).
\end{equation}
Now, recall Jacobi's rule for the derivative of a determinant, 
\begin{equation}
\frac{\partial h(\lambda)}{\partial \lambda} = h \, h^{-1}_{ij}  \frac{\partial h_{ji}}{\partial \lambda}, 
\end{equation}
where $\lambda$ is an arbitrary parameter on which the elements of ${\bf h}$ depend.
This allows us to write
\begin{equation}
\frac{\partial } {\partial t} \frac{1}{\sqrt{h}} \Big|_{\bf r} =  -\frac{1}{2 \sqrt{h}} 
  \left( \frac{\partial \xi_i } {\partial r_\lambda}  \frac{\partial^2 r_\lambda } {\partial \xi_i \partial t} - 
  \frac{\partial \xi_i}{\partial r_j}  \frac{\partial^2 r_j }{\partial \xi_i \partial \xi_\beta}
  \frac{\partial \xi_\beta}{\partial r_\lambda} \frac{\partial r_\lambda}{\partial t} \right).
\end{equation}
At this point, observe that for a matrix ${\bf A}$ that depends parametrically on $\lambda$,
we have
\begin{equation}
\frac{\partial {\bf A}^{-1}(\lambda)}{\partial \lambda} =
-{\bf A}^{-1}   \frac{\partial {\bf A}(\lambda)}{\partial \lambda} {\bf A}^{-1}
\end{equation}
We use this relationship to observe that
\begin{equation}
 -\frac{\partial \xi_i}{\partial r_j}  \frac{\partial^2 r_j }{\partial \xi_i \partial \xi_\beta}
  \frac{\partial \xi_\beta}{\partial r_\lambda} = 
  \frac{\partial }{\partial \xi_i} \left( \frac{\partial \xi_i}{\partial r_\lambda} \right).
\end{equation}  
This allows us to write the derivative of the volume factor in a compact form,
\begin{equation}
\frac{\partial } {\partial t} \frac{1}{\sqrt{h}} \Big|_{\bf r} =  -\frac{1}{2 \sqrt{h}} 
  \frac{\partial }{\partial \xi_i} \left( \frac{\partial \xi_i}{\partial r_\lambda} \frac{\partial r_\lambda}{\partial t}  \right).
\end{equation}
By using the quantities that we introduced earlier, we can equivalently write
\begin{equation}
\frac{\partial } {\partial t} \frac{1}{\sqrt{h}} \Big|_{\bf r} =  -\frac{1}{2 \sqrt{h}} 
  \frac{\partial }{\partial \xi_\beta} \left(  g^{-1}_{\beta \gamma} A_\gamma \right).
\end{equation}
%
%
%
%
%
%
After few straightforward steps of algebra, one finally 
arrives at 
\begin{equation}
i \frac{\partial \psi}{\partial t} \Big|_{\bf r} =
  \frac{i}{\sqrt{h}}  \frac{\partial \cur{\psi}}{\partial t} \Big|_{\bm{\xi}}
   + \frac{1}{2 \sqrt{h}} \left( A_\beta g^{-1}_{\beta \gamma} \cop{p}_\gamma
   + \cop{p}_\beta g^{-1}_{\beta \gamma} A_\gamma \right) \cur{\psi}.
\mylabel{tder}
\end{equation}
Then, by observing that the effective scalar potential of Eq.~(\ref{phidef})
can also be written as
\begin{equation}
\phi = A_\beta g^{-1}_{\beta \gamma} A_\gamma,
\end{equation}
one can combine Eq.~(\ref{tder}) with the kinetic terms that we have derived
in the previous subsection, leading to Eq.~(\ref{schr-xi}).
%

\section{Static perturbation in the linear regime}

\mylabel{app:b}

In this Appendix we shall provide an explicit expression for 
the ``static'' perturbation of Sec.~\ref{sec:staticpert}, in the
specific case of a monochromatic perturbation at wavevector ${\bf q}$.
We shall also show that it reduces to Hamann's metric perturbation
at first order in ${\bf q}$.

The first-order Hamiltonian can be decomposed as follows,
\begin{equation}
\hat{\mathcal{H}}_{\bf k,q}^{(\lambda_\beta)} = \hat{\mathcal{T}}_{\bf k,q}^{(\lambda_\beta)} + \hat{V}_{\rm geom,\bf q}^{(\lambda_\beta)} +
\hat{V}_{\rm H,\bf q}^{(\lambda_\beta)} + \hat{V}_{\rm XC,\bf q}^{(\lambda_\beta)},
\mylabel{Hstat}
\end{equation}
where the four terms on the rhs are related, respectively, to the kinetic ($\hat{\mathcal{T}}$) operator, and the 
geometric, Hartree and exchange-correlation potentials. 
In the following, we shall 
discard all the dynamical terms that emerge from the time
derivative.
Note that all the operators in Eq.~(\ref{Hstat}) are cell-periodic, i.e.,
the static curvilinear-space Hamiltonian can be written, in the linear regime, as
\begin{equation}
\cop{\mathcal{H}}(\bm{\lambda}) = \hat{\mathcal{H}}^{(0)} + \lambda_\beta e^{i{\bf q} \cdot {\bf r}} \hat{\mathcal{H}}_{\bf k,q}^{(\lambda_\beta)} + \cdots.
\end{equation}
Note the use of ${\bf r}$ instead of $\bm{\xi}$ to indicate the coordinates in curvilinear space;
we shall follow this convention henceforth (and omit the tilde on the curvilinear operators).
For a generic perturbation $\mathcal{O}$, we shall also use the following notation convention
to distinguish the full operator from its cell-periodic part,
\begin{equation}
\hat{\mathcal{O}}_{\bf k}^{(\lambda_\beta)}({\bf q}) = e^{i{\bf q} \cdot {\bf r}} \hat{\mathcal{O}}_{\bf k,q}^{(\lambda_\beta)}.
\end{equation}

\subsection{Kinetic term}

The curvilinear kinetic operator can be expanded, in powers of the deformation amplitude, as
\begin{equation}
\hat{\mathcal{T}}_{{\bf k}} = \frac{\hat{p}_{{\bf k}}^2 }{2} - \hat{p}_{{\bf k} \beta} \, \varepsilon_{\beta \gamma}({\bf r}) \, \hat{p}_{{\bf k} \gamma} + \cdots,
\end{equation}
where $\varepsilon_{\beta \gamma}({\bf r})$ is the symmetric strain tensor associated with a
generic inhomogeneous deformation.
In the specific case of a monochromatic displacement wave, the strain reads as
\begin{equation}
\varepsilon_{\beta \gamma}({\bf r}) = \frac{i}{2} \left( \lambda_\beta q_\gamma + \lambda_\gamma q_\beta  \right) e^{i {\bf q} \cdot {\bf r} }.
\end{equation}
This immediately leads to 
\begin{equation}
\hat{\mathcal{T}}_{\bf k,q}^{(\lambda_\beta)} = -\frac{i}{2} \left[  (\hat{p}_{{\bf k}\beta} +  q_\beta   ) \, {\bf q} \cdot  \hat{\bf p}_{{\bf k}}     +  
                                                                     (\hat{\bf p}_{{\bf k}} +  {\bf q}   ) \cdot  {\bf q} \, \hat{p}_{{\bf k}\beta}   \right].
\end{equation}
At first order in ${\bf q}$, we have
\begin{equation}
\frac{\partial \hat{\mathcal{T}}_{\bf k,q}^{(\lambda_\beta)}}{\partial q_\gamma}\Big|_{{\bf q}=0} = -i \, \hat{p}_{{\bf k}\beta} \hat{p}_{{\bf k}\gamma} =
    i \hat{\mathcal{T}}_{\bf k}^{(\beta\gamma)},
\end{equation}
where we have indicated with a superscript $(\beta \gamma)$ the response to a uniform 
strain, $\varepsilon_{\beta \gamma}$, within Hamann's formalism. 
%

\subsection{Geometric potential}

By retaining only terms that are linear in the deformation amplitude, we have
\begin{equation}
V_{\rm geom} = \frac{1}{2} \partial_\alpha \mathcal{A}_\alpha, \qquad
\mathcal{A}_\alpha = -\frac{1}{2}  \bm{\lambda} \cdot {\bf q} \, q_\alpha e^{i \bm{\xi} \cdot {\bf q} }.
\end{equation}
Then, one immediately obtains  
\begin{equation}
V^{(\beta,{\bf q})}_{\rm geom} = -\frac{i}{4} q_\beta \, q^2.
\end{equation}
This structureless potential is irrelevant for either the uniform strain
or the strain-gradient response, as it is of third order in ${\bf q}$.

\subsection{Electrostatic potential}

Recall Poisson's equation in curvilinear space,
\begin{equation}
\partial_\alpha (h g^{\alpha \gamma}  \partial_\gamma V_{\rm H}) = -4 \pi ({n}_{\rm el} - {\rho}_{\rm ion}).
\end{equation}
In the linear limit one has
\begin{eqnarray}
h g^{-1}_{\alpha \gamma} &=&  \delta_{\alpha \gamma} + i (\bm{\lambda} \cdot {\bf q} \delta_{\alpha \gamma} - \lambda_\alpha q_\gamma - \lambda_\gamma q_\alpha ) e^{i {\bf r} \cdot {\bf q} }, \\
V_{\rm H} &=& V_{\rm H}^{(0)} + \lambda_\beta e^{i {\bf q} \cdot {\bf r} } V_{\rm H,{\bf q}}^{(\lambda_\beta)}, \\
{n}_{\rm el} &=& {n}_{\rm el}^{(0)} + \lambda_\beta e^{i {\bf q} \cdot {\bf r} } n_{\rm el,{\bf q}}^{(\lambda_\beta)}.
\end{eqnarray}
(Note that the ionic point charges do not move within the curvilinear frame; thus, their
density, ${\rho}_{\rm ion}$, is unsensitive to the deformation.)
By collecting the terms that are linear in $\lambda_\beta$, and by following analogous
derivation steps as in Ref.~\onlinecite{artgr}, we have
\begin{equation}
 |\bm{\nabla} + i {\bf q}|^2  V_{\rm H,{\bf q}}^{(\lambda_\beta)} =
 - 4 \pi \left( n_{\rm el,{\bf q}}^{(\lambda_\beta)} + n_{\rm met,{\bf q}}^{(\lambda_\beta)} \right),
\mylabel{poisson-q}
\end{equation}
where the ``metric density'' $n_{\rm met,{\bf q}}^{(\lambda_\beta)}$ is given in terms
of the ground-state Hartree potential,
\begin{equation}
n_{\rm met,{\bf q}}^{(\lambda_\beta)} = \frac{i}{4\pi}  (\partial_\alpha  + i q_\alpha)  
 [ \delta_{\alpha \gamma} q_\beta - \delta_{\alpha\beta } q_\gamma 
        -  \delta_{\beta \gamma} q_\alpha ] \partial_\gamma V_{\rm H}^{(0)}.
\end{equation}        
At order zero in ${\bf q}$, all the scalar fields involved in Eq.~(\ref{poisson-q})
manifestly vanish. (The kinetic and geometric perturbations discussed in the previous subsections 
both vanish, yielding a null first-order density; $n_{\rm met,{\bf q}}^{(\lambda_\beta)}$ 
vanish as well, as it has a leading dependence on ${\bf q}$; the first-order potential then
vanishes as well as a consequence of Eq.~(\ref{poisson-q}).)
At first order in ${\bf q}$, one has
\begin{equation}
 {\nabla}^2  V_{\rm H,\zeta}^{(\lambda_\beta)} =
 - 4 \pi \left( n_{\rm el,{\zeta}}^{(\lambda_\beta)} + n_{\rm met,{\zeta}}^{(\lambda_\beta)} \right),
\end{equation}
where we have indicated the derivatives with respect to $q_\zeta$
calculated at ${\bf q}=0$ with a $\zeta$ subscript, and
\begin{equation}
n_{\rm met,{\zeta}}^{(\lambda_\beta)} = \frac{i}{4\pi} \left( \nabla^2 \delta_{\zeta \beta} - 2 \partial_\zeta \partial_\beta \right) V_{\rm H}^{(0)}.
\end{equation}
Finally, by expressing the cell-periodic scalar fields in Fourier space, 
we obtain
\begin{equation}
V_{\rm H,\gamma}^{(\lambda_\beta)}  = \frac{4\pi}{G^2}    \left[
    n_{\rm el,{\gamma}}^{(\lambda_\beta)} 
   - i n^{(0)} \left(  \delta_{\beta \gamma} - 2 \frac{ G_\beta G_\gamma  }{G^2} \right)  
   \right],
\end{equation}
where $n^{(0)} = n_{\rm el}^{(0)} - \rho_{\rm ion}$ is the ground-state electronic density
minus the ionic point-charges (i.e., it corresponds to the opposite of the total 
charge density of the crystal).
After observing that $n_{\rm el,{\gamma}}^{(\lambda_\beta)} = i n_{\rm el}^{(\beta \gamma)}$,  
one can easily verify that the above formula coincides (modulo a factor of $i$) with 
Hamann's Eq.~(57).

\subsection{XC potential}

Starting from Eq.~(\ref{exc}), one can write the exchange-correlation potential as
\begin{equation}
V_{\rm XC}(\bm{\xi}) = \frac{\delta E_{\rm XC}}{\delta n (\bm{\xi})} = \epsilon_{\rm XC}(h^{-1} n) + h^{-1} n  \epsilon'_{\rm XC}(h^{-1} n).
\end{equation}
(The prime symbol indicates a first derivative with respect to the particle density.)
After a few algebra steps, one arrives at an expression for the perturbed potential,
\begin{equation}
 V_{\rm XC,{\bf q}}^{(\lambda_\beta)} ({\bf r}) = K_{\rm XC} ({\bf r}) \, \left[ n_{\rm el,{\bf q}}^{(\lambda_\beta)}({\bf r}) - i q_\beta n_{\rm el}^{(0)}({\bf r})\right],  
\end{equation}
where
\begin{equation}
K_{\rm XC} = 2 \epsilon'_{\rm XC}( n^{(0)}) + n^{(0)}  \epsilon''_{\rm XC}(n^{(0)})
\end{equation}
is the exchange-correlation kernel, and the contribution that
depends on $n^{(0)}$ originates from the derivative of
the inverse determinant,
\begin{equation}
h^{-1} = 1 - i \bm{\lambda} \cdot {\bf q} e^{i \bm{\xi} \cdot {\bf q} }.
\end{equation}
Again, the first-order potential vanishes at ${\bf q}=0$ and coincides with
Hamann's metric formulation of the uniform strain perturbation at first 
order in ${\bf q}$.

\bibliography{merged}

\end{document}